\documentclass[aps,pra,twocolumn,superscriptaddress,showpacs,amsmath,amsfonts,amssymb]{revtex4-1}
\usepackage{amsmath,amsfonts,amssymb,color}
\usepackage{amsthm}
\usepackage{leftidx}
\usepackage{graphicx}
\usepackage{xcolor}
\usepackage{dcolumn}
\usepackage{bm}
\usepackage{epstopdf}
\usepackage{epsfig}

\newcommand{\JB}{\textcolor{black}}
\newcommand{\R}{\textcolor{black}}
\newcommand{\MS}{\textcolor{black}}

\begin{document}
	\title{Topological and dynamical features of periodically driven spin ladders }
	\author{Raditya Weda Bomantara}
	\email{raditya.bomantara@sydney.edu.au}
	\affiliation{%
		Centre for Engineered Quantum Systems, School of Physics,
		University of Sydney, Sydney, New South Wales 2006, Australia
	}
	\author{Sen Mu}
	\email{senmu@u.nus.edu}
	\affiliation{%
		Department of Physics, National University of Singapore, Singapore 117543
	}
	\author{Jiangbin Gong}%
	\email{phygj@nus.edu.sg}
	\affiliation{%
		Department of Physics, National University of Singapore, Singapore 117543
	}
	\date{\today}
	
	
	\vspace{2cm}
	
	\begin{abstract}
\JB{Studies of periodically driven one-dimensional many-body systems have advanced our understanding of complex systems and stimulated promising developments in quantum simulation. It is hence of interest to go one step further, by investigating the topological and dynamical aspects of periodically driven spin ladders as clean quasi-one-dimensional systems with spin-spin interaction in the rung direction.} \R{Specifically, we find that such systems display subharmonic magnetization dynamics reminiscent to that of discrete time crystals (DTCs) at finite system sizes. Through the use of generalized Jordan-Wigner transformation, this feature can be attributed to presence of corner Majorana $\pi$ modes (MPMs), which are of topological origin, in the systems' equivalent Majorana lattice. Special emphasis is placed on how the coupling in the rung direction of the ladder prevents degeneracy from occurring between states differing by a single spin excitation, thus preserving the MPM-induced $\pi/T$ quasienergy spacing of the Floquet eigenstates in the presence of parameter imperfection.  This feature, which is absent in their strict one-dimensional counterparts, may yield fascinating consequences in future studies of higher dimensional Floquet many-body systems.}
	\end{abstract}
	
	\maketitle
	
	\section{Introduction}
	
\JB{In recent years, periodically driven many-body systems have attracted vast experimental and theoretical efforts.  Ergodicity-breaking phenomena in such systems, especially the concept of discrete time crystal (DTC) \cite{DTC1,DTC2,DTC3,DTC4,DTC5,DTC6,DTC7,DTC8,DTC9,DTC10,DTC11,DTC12,DTC13,DTC14,DTC15,DTC16,DTC17,DTC18,DTC19,DTC20,DTC21,DTC22,DTC23,DTC24,DTC25,DTC26,DTC27,DTC29,DTC30,DTC31,DTC32} have opened up new research avenues for
quantum simulation \cite{DTCqs,DTCqs2}, quantum computation \cite{DTCqc},  and studies of other condensed matter phenomena in the time-domain \cite{DTCcm1,DTCcm2,DTCcm3,DTCcm5,DTCcm6}.  Indeed, owing to the complexity of driven many-body systems,}  \R{theoretical and experimental advances} \JB{may potentially lead to vast opportunities towards quantum supremacy, where experimental demonstrations or findings can be beyond the capability of any classical computer.}

\JB{The nonequilibirum nature of periodically driven systems can also be exploited to access or generate a variety of topological phases of matters, many of which are not available in equilibrium systems. In particular, studies of non-equilibrium Floquet topological matter \cite{FMF1,FMF2,FMF3,DTCqc,FMF5,FMF6,FMF7,FMF8} have revealed that symmetry-protected topological excitation in periodically many-body systems with period $T$ can have $\pi/T$ quasienergy excitations on top of zero quasienergy excitations.} \R{The topologically protected $\pi/T$ quasienergy excitation are of particular interest because they result in all Floquet eigenstates coming in pairs with $\pi/T$ quasienergy separation. As an immediate consequence, a linear superposition of two $\pi/T$ quasienergy separated eigenstates can be formed, which directly breaks the discrete time translational symmetry and leads to subharmonic oscillations in the system dynamics. This simple physical picture emphasizes the importance of investigating the topological and the spectral (hence also the dynamical) properties together to explore interesting many-body physics. }   

\JB{Ongoing studies of periodically driven many-body systems are mostly based on one-dimensional systems. For instance, the suppression of thermalization via many-body localization (MBL) \cite{MBL1,MBL2,MBL3,MBL4} is known to be important in one-dimensional (1D) DTCs, but how MBL works in higher dimensions is largely unknown.  This perhaps explains why best known examples of DTCs is established in 1D setting.  Developing more theoretical insights  into periodically driven many-body systems in higher-dimensions is thus important from theoretical point of view and also necessary to guide future experimental designs.} 

\R{As a first step towards understanding higher-dimensional Floquet many-body systems, this work aims to reveal some important physics of periodically driven spin ladders at moderate system sizes. Special emphasis is placed on the subharmonic magnetization response to the periodic driving.} \JB{Because we study clean systems here, such subharmonic response cannot be taken as a genuine signature of DTCs in the strict sense.} \R{Nevertheless, at finite system sizes, this dynamical feature simulates that of a true DTC over a considerable time scale. Studying subharmonic magnetization dynamics} \JB{also provides an important angle to digest how spin-spin coupling in the rung direction of the ladder impacts on the spectral properties of the system.  Referring to a terminology from studies of non-equilibrium Floquet topological matter \cite{FMF1,FMF2,FMF3,DTCqc,FMF5,FMF6,FMF7,FMF8}, the periodically driven spin ladders can accommodate the so-called Majorana $\pi$ modes (MPMs), i.e.,  mutually anticommuting Hermitian operators that also anticommute with the system's one-period (Floquet) propagator. Specifically, by rewriting the system Hamiltonian specified below in its equivalent Majorana lattice model,  these MPMs correspond to, \JB{at least for special parameter values}, the topologically protected Majorana modes at the two lattice corners, in the spirit of second-order topological phases \cite{HTI-1,HTI0,HTI1,HTI2,HTI3,HTI4,HTI5,HTI6,HTI7,HTI8,HTI9,HTI10,HTI11,HTI12,HTI12b,HTI12c,HTI13,HTI14,HTI15,HTI16,HTI17,HTI18,HTI19,HTI20,HTI21,HTI22,HTI23,HTI24,HTI25,HTI26,FHTI1,FHTI2,FHTI3,FHTI4,FHTI5,FHTI6,FHTI7,FHTI8,FHTI9}. With this understanding, the spectral properties of a spin ladder at finite system sizes can be investigated in terms of the} \R{fate of these corner MPMs at various system sizes and parameters.}

	This paper is organized as follows. In Sec.~II, we present our model and elucidate the role of MPMs in yielding subharmonic dynamics. In Sec.~III, we compare the proposed model with its 1D counterpart in terms of their Floquet eigenstates structure and computationally investigate its implication on the robustness of the $2T$ magnetization dynamics against variation in the system size and choice of initial states. Finally, we conclude this paper, briefly discuss the potential experimental realization of the proposed model, and present avenues for future work in Sec.~IV. 
	
	\section{Periodically driven spin ladder}
	Consider a quasi-one-dimensional rectangular lattice of size $N_x\times N_y$ subjected to the time-periodic Hamiltonian 
	\begin{widetext}
		\begin{equation}
		H(t) = \begin{cases}
		-\sum_{\langle i,j \rangle} \left(J_x \sigma_{z,i,j} \sigma_{z,i+1,j} + J_y\sigma_{z,i,j+1} \sigma_{z,i,j}\right) & \text{ for } nT<t\leq (n+1/2)T \\
		\sum_{i,j} h\sigma_{x,i,j} & \text{ for } (n+1/2)T<t\leq (n+1)T
		\end{cases} \;, \label{sys}
		\end{equation}
	\end{widetext} 
	where $\left\lbrace\sigma_{s,i,j}:s=x,y,z\right\rbrace$ is the set of three Pauli matrices acting at lattice site $(i,j)$, $J_x$ and $J_y$ are the ZZ interactions in the $x$- and $y$-directions respectively, $h$ is the Zeeman field strength, $T$ is the driving period, and $n\in \mathbb{Z}$. While Eq.~(\ref{sys}) can generally be interpreted as a two-dimensional (2D) system, we choose to term it a spin ladder for the following reasons: (i) we work in the regime $J_x\ll J_y$ and (ii) we fix $N_y=2$ in the most case studied in this paper.
	
	As a time-periodic Hamiltonian, Eq.~(\ref{sys}) can be characterized in terms of its one-period time evolution operator (to be called \emph{Floquet operator} onwards) \cite{Flo1,Flo2}
	\begin{equation}
	U \equiv U(T;0)=\mathcal{T} \exp\left(-\mathrm{i} \int_0^T H(t) dt \right)\;,
	\end{equation}
	where $\mathcal{T}$ is the time-ordering operator and $\hbar=1$ is taken here and in the remainder of this paper. The system's quasienergy $\varepsilon \in (-\pi/T,\pi/T]$ and its associated Floquet eigenstate $|\varepsilon\rangle$ are defined from the eigenvalues $(\exp\left[-\mathrm{i}\varepsilon T\right])$ and eigenstates of $U$ respectively. A quasienergy is only defined modulo $2\pi/T$, so that $\varepsilon$ and $\varepsilon+2\pi\ell/T$ with $\ell\in\mathbb{Z}$ represent the same physics.

The Floquet operator associated with Eq.~(\ref{sys}) can be easily obtained as
 \begin{eqnarray}
U &=& e^{-\mathrm{i} \sum_{i,j} \frac{hT}{2} \sigma_{x,i,j}}  \nonumber \\
&& \times e^{\left(\sum_{i,j} \mathrm{i} \left[\frac{J_yT}{2}\sigma_{z,i,j+1}\sigma_{z,i,j}+\frac{J_xT}{2} \sigma_{z,i+1,j}\sigma_{z,i,j}\right]\right)} \;, \label{Flospin} 
\end{eqnarray}
By employing the generalized JW transformation detailed in Appendix~\ref{app1}, the different spin operators appearing in Eq.~(\ref{Flospin}) can be written in terms of mutually anticommuting Majorana operators $\gamma_{A,i,j}$ and $\gamma_{B,i,j}$ as 
\begin{eqnarray}
\sigma_{x,i,j} &=& \mathrm{i} \gamma_{A,i,j} \gamma_{B,i,j} \;, \sigma_{z,i,j+1}\sigma_{z,i,j} = \mathrm{i} \gamma_{B,i,j} \gamma_{A,i,j+1} \;, \nonumber \\
\sigma_{z,i+1,j}\sigma_{z,i,j} &=& \mathrm{i} \gamma_{B,i,j} \gamma_{A,i+1,j} \prod_{j<n\leq N_y} \left(\gamma_{A,i,n} \gamma_{B,i,n}\right) \nonumber \\
&& \times \prod_{m<j} \left(\gamma_{A,i+1,m} \gamma_{B,i+1,m}\right) \;. \label{opmap} 
\end{eqnarray}
Figure 1 illustrates some representative spin operators above when $\gamma_{A,i,j}$ and $\gamma_{B,i,j}$ are arranged into a
2D square lattice. The highly nonlocal terms arising from $\sigma_{z,i+1,j}\sigma_{z,i,j}$ are unavoidable \cite{Chapman} and highlight the strongly interacting nature of our system. 
\begin{center} 
	\begin{figure}
		\includegraphics[scale=0.4]{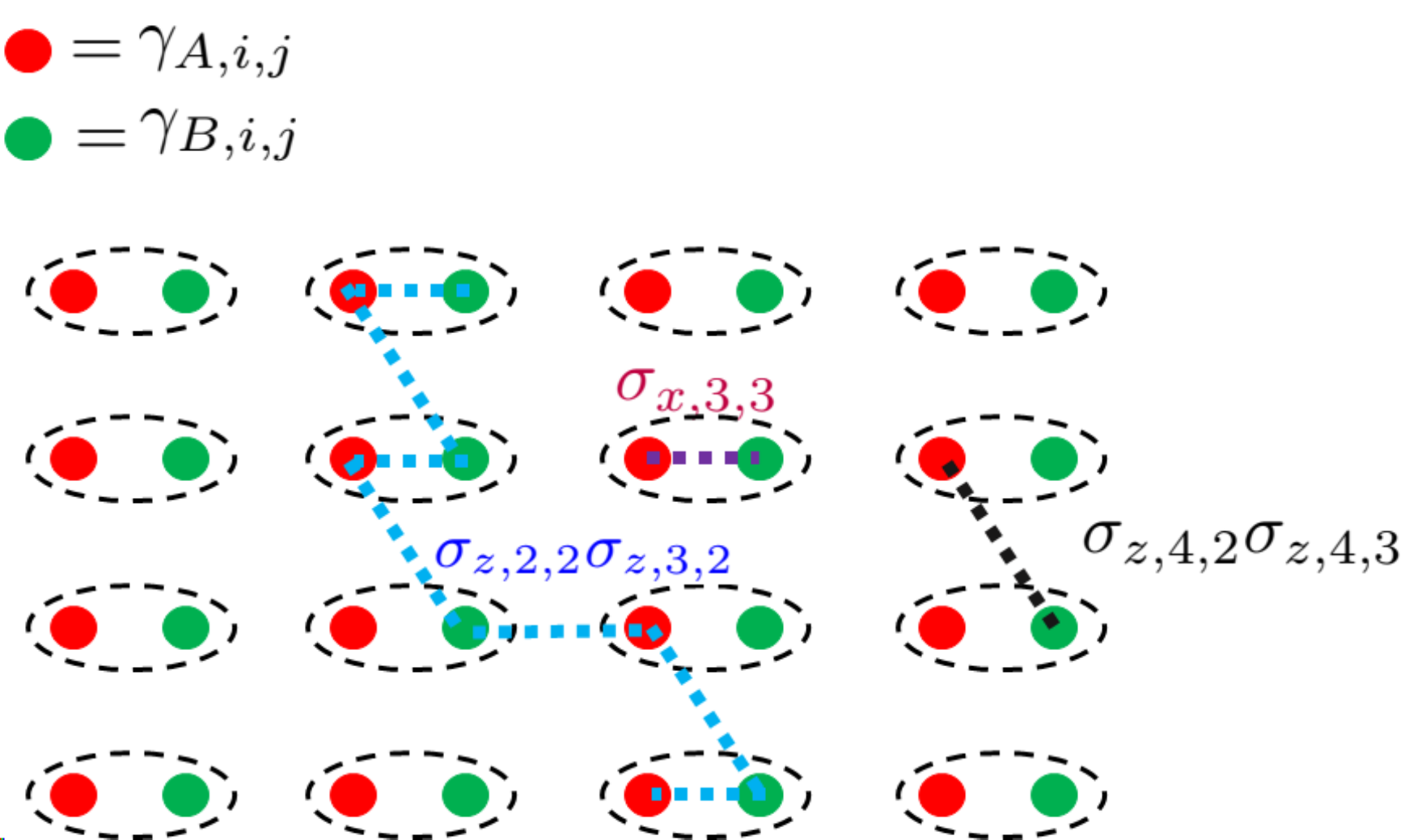}
		\caption{Schematic of various terms in Eqs.~(\ref{sys}) and (\ref{Flospin}) when written in terms of Majorana operators. There, each term is a product of all Majorana operators contained in its respective dotted line.}
		\label{pic1}
	\end{figure}
\end{center}

\subsection{MPMs as topological quasienergy excitations} 

Note first that at $N_x=1$, Eq.~(\ref{sys}) reduces to a clean (disorder-free) variation \cite{DTC8} of the  spin-based 1D DTC models \cite{DTC2,DTC5} that is analytically solvable,
with the presence or absence of Majorana zero modes (MZMs) and MPMs yielding the phase diagram in Fig.~\ref{result1}(a) (See Appendix~\ref{app1} for its analytical derivation).  Each regime in the phase diagram is labeled by the four phases identified in Ref.~\cite{DTC5}. Specifically, regardless of the boundary conditions, the paramagnet (PM), $0$-spin glass ($0$-SG), and $\pi$-spin glass ($\pi$-SG) host, respectively, neither MPM nor MZM, a pair of MZMs, and a pair of MPMs. On the other hand, in the $0\pi$-paramagnet regime, the system hosts a pair of MZMs and a pair of MPMs only under open boundary conditions (OBC); otherwise, it does not support any MZM or MPM. Outside $(h,J_y)\in [0,\pi/2]\times [0,\pi/2]$ (in $T/2=1$ units), the four phases above repeat themselves, e.g., $\pi$-SG phase repeats up to $h=\pi$ at $J_y=\pi/4$, followed by another $0$-SG phase as $h$ further increases.
\begin{center} 
	\begin{figure*}
		\includegraphics[scale=1.0]{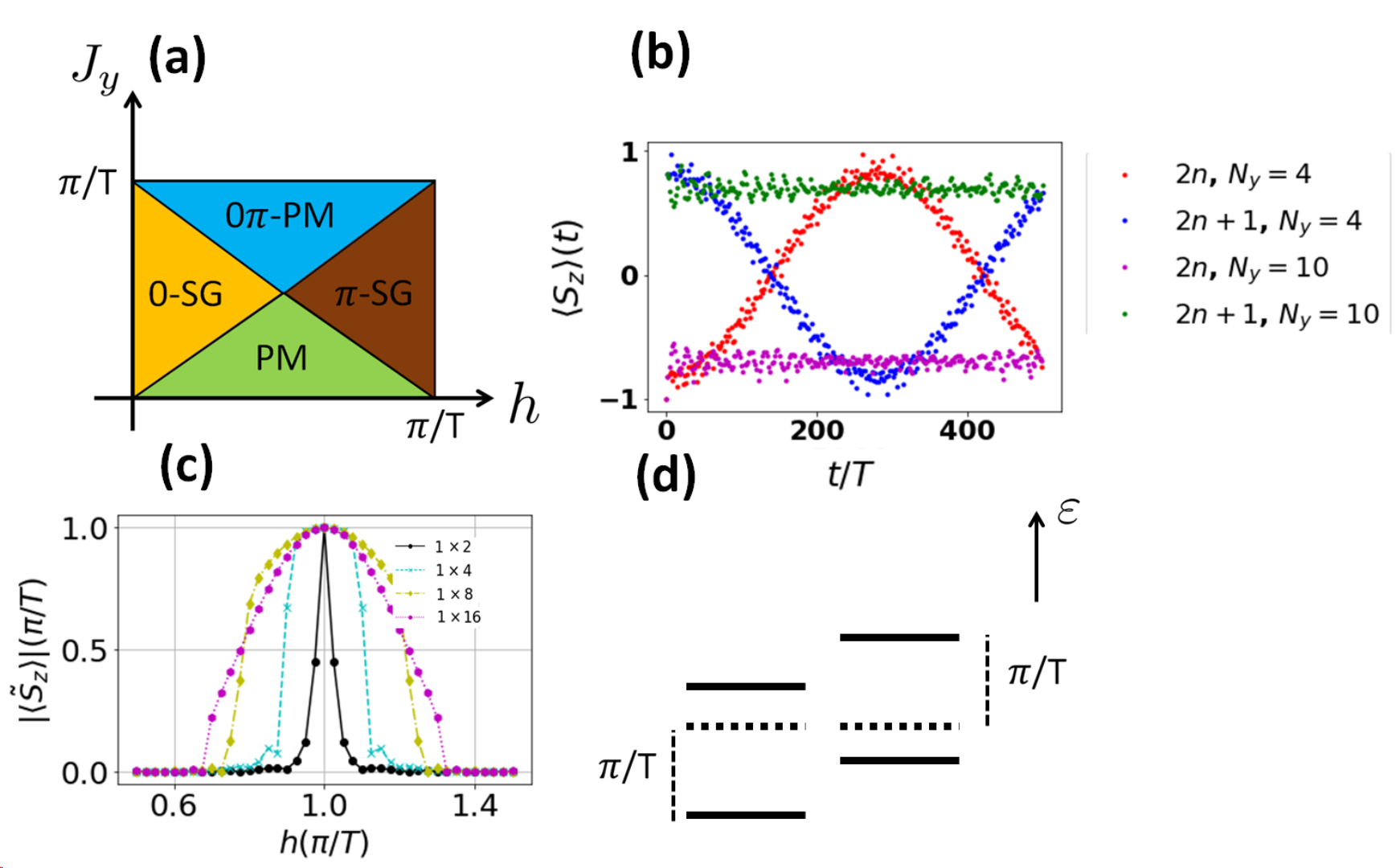}
		\caption{(a) Phase diagram of the periodically driven Ising spin chain Eq.~(\ref{sys}) in the 1D limit, i.e., $N_x=1$. (b) Stroboscopic magnetization dynamics at $N_x=1$, $J_y=0.6\pi/T$, and $h=0.8\pi/T$. (c) The subharmonic response of the system's average magnetization at $\Omega=\pi/T$, $J_x=0.05\pi/T$, $J_y=0.6\pi/T$, and various lattice configurations. There, the system is initialized in a product state $|\uparrow\cdots \uparrow \rangle$ and the average magnetization is evolved up to $2000T$. (d) Perturbation-induced degeneracy splitting may break the $\pi/T$ quasienergy spacing structure.}
		\label{result1}
	\end{figure*}
\end{center}

\JB{Since the case of $N_x=1$ can be exactly solved, it is straightforward to digest} the role of MPMs in partially establishing two main dynamical features analogous to those of DTCs \cite{DTC2,DTC3,DTC8}: i) the existence of $2T$-periodic state and ii) its robustness against variation in the system parameters. To this end, we first define a \emph{quasienergy excitation} operator $\gamma_\xi$ satisfying $U\gamma_\xi U^\dagger=\exp\left(-\mathrm{i} \xi\right)\gamma_\xi $. As its name suggests, it ``excites" a given quasienergy eigenstate $|\varepsilon\rangle$ to another $|\varepsilon+\xi\rangle=\gamma_\xi |\varepsilon \rangle $ with quasienergy $\varepsilon+\xi$. In particular, MPMs are defined as quasienergy $\pi/T$ excitations, whose existence implies the presence of two quasienergies $\varepsilon$ and $\varepsilon+\pi/T$ in the system. In this case, a superposition state $|\varepsilon\rangle +|\varepsilon+\pi/T\rangle$ returns to itself upon applying $U$ twice (is $2T$ periodic), thus establishing the first feature above. It is also worthwhile noting that, as further discussed below, each $|\varepsilon\rangle$ and $|\varepsilon+\pi/T\rangle$ ideally takes the form of a symmetric/anti-symmetric cat state. \R{Consequently, an equal weight superposition of $|\varepsilon\rangle$ and $|\varepsilon+\pi/T\rangle$, which demonstrates $2T$ periodicity, yields a product state. Such a product state, and thereby its associated subharmonic dynamics, is naturally robust against decoherence mechanism that tends to break entanglements.} 

Second, the presence of MPMs in the above model can be traced back from the existence of \emph{topological edge modes} arising in their equivalent Majorana fermionic description via a Jordan-Wigner (JW) transformation or its generalization (see Eq.~(A1) in Appendix~\ref{app1}). Consequently, such MPMs are not a result of fine tuning and exist for a range of system parameters, thereby preserving the $2T$ periodic state identified above. Moreover, due to the nonlocal nature of the JW-like transformation, the original spin system can always be mapped to its corresponding Majorana fermionic description under OBC, possibly with additional nonlocal terms which may not always lead to the pairwaise annihilation of MPMs. This in turn yields the $\pi$-SG regime scenario in which MPMs exist \emph{regardless} of the boundary conditions applied to the original spin system.

\subsection{Robustness of corner MPMs}

At $N_x>1$ and $hT/2=\pi/2$, a quick inspection of Fig.~\ref{pic1} allows the identification of $\gamma_{A,1,1}=\sigma_{z,1,1}$ and $\gamma_{B,N_x,N_y}=\left(\prod_{(l<N_x),(m\leq N_y)} \sigma_{x,l,m} \right) \left(\prod_{(l<N_y)} \sigma_{x,N_x,l} \right) \sigma_{y,N_x,N_y}$ as the system's two MPMs.  {Note that in the equivalent Majorana description of our system, $\gamma_{A,1,1}$ and $\gamma_{B,N_x,N_y}$ are precisely located at the two corners (see Fig.~\ref{pic1}), suggesting that they may share a similar origin as the topological corner modes in typical Floquet second-order topological superconductors \cite{FHTI8,FHTI9}}. 

\JB{In general conditions, cases with $N_x\ne 1$} \R{are no longer integrable.} \JB{This is precisely why the present study for limited system sizes is still necessary and useful.}  Though the fate of the above identified corner MPMs is no longer analytically tractable at $h\neq \pi/2$, we may still numerically probe their existence by defining and evaluating the two corner-state spectral functions at zero and $\pi/T$ quasienergy, i.e.,
\begin{eqnarray}
s_{0,1} &=& \mathcal{N}_1  \sum_{n\in \mathcal{X}} \int_{-\epsilon}^{\epsilon} S_1(\varepsilon_n, \eta) d\eta  \;, \nonumber \\
s_{0,2} &=& \mathcal{N}_2 \sum_{n\in \mathcal{X}} \int_{-\epsilon}^{\epsilon} S_2(\varepsilon_n, \eta) d\eta \;, \nonumber \\
s_{\pi,1} &=& \mathcal{N}_1 \sum_{n\in \mathcal{X}} \left( \int_{-\pi/T-\epsilon}^{-\pi/T+ \epsilon} S_1(\varepsilon_n, \eta) d\eta + \int_{\pi/T-\epsilon}^{\pi/T+ \epsilon} S_1(\varepsilon_n, \eta)  d\eta \right) \nonumber \;, \\
s_{\pi,2} &=& \mathcal{N}_2 \sum_{n\in \mathcal{X}} \left( \int_{-\pi/T-\epsilon}^{-\pi/T+ \epsilon} S_2(\varepsilon_n, \eta) d\eta + \int_{\pi/T-\epsilon}^{\pi/T+ \epsilon} S_2(\varepsilon_n, \eta)  d\eta \right) \nonumber \;,
\end{eqnarray}
where $\epsilon\ll 1$, $\mathcal{X}$ represents a rather arbitrarily chosen integer between $1$ and the size of the system's Hilbert space, $\mathcal{N}_1$ and $\mathcal{N}_2$ are normalization constants such that $\sum_{n\in \mathcal{X}} \int_{-\pi/T}^{\pi/T} S_1(\varepsilon_n,\eta) d\eta = \sum_{n\in \mathcal{X}} \int_{-\pi/T}^{\pi/T} S_2(\varepsilon_n,\eta) d\eta =1$, and
\begin{eqnarray}
S_1(\varepsilon_n, \eta) &=& \sum_{\varepsilon_m} \delta(\varepsilon_n-\varepsilon_m -\eta) |\langle \varepsilon_n | \gamma_{A,1,1} | \varepsilon_m \rangle |^2 \;, \nonumber \\    
S_2(\varepsilon_n, \eta) &=& \sum_{\varepsilon_m} \delta(\varepsilon_n-\varepsilon_m -\eta) |\langle \varepsilon_n | \gamma_{B,N_x,N_y} | \varepsilon_m \rangle |^2 \;.
\end{eqnarray}
In particular, the quantities $s_{0,1}$, $s_{0,2}$, $s_{\pi,1}$, and $s_{\pi,2}$ are adapted from the edge spectral functions defined in Ref.~\cite{pzmandppm} and can be understood as follows.

\begin{center} 
	\begin{figure}
		\includegraphics[scale=0.51]{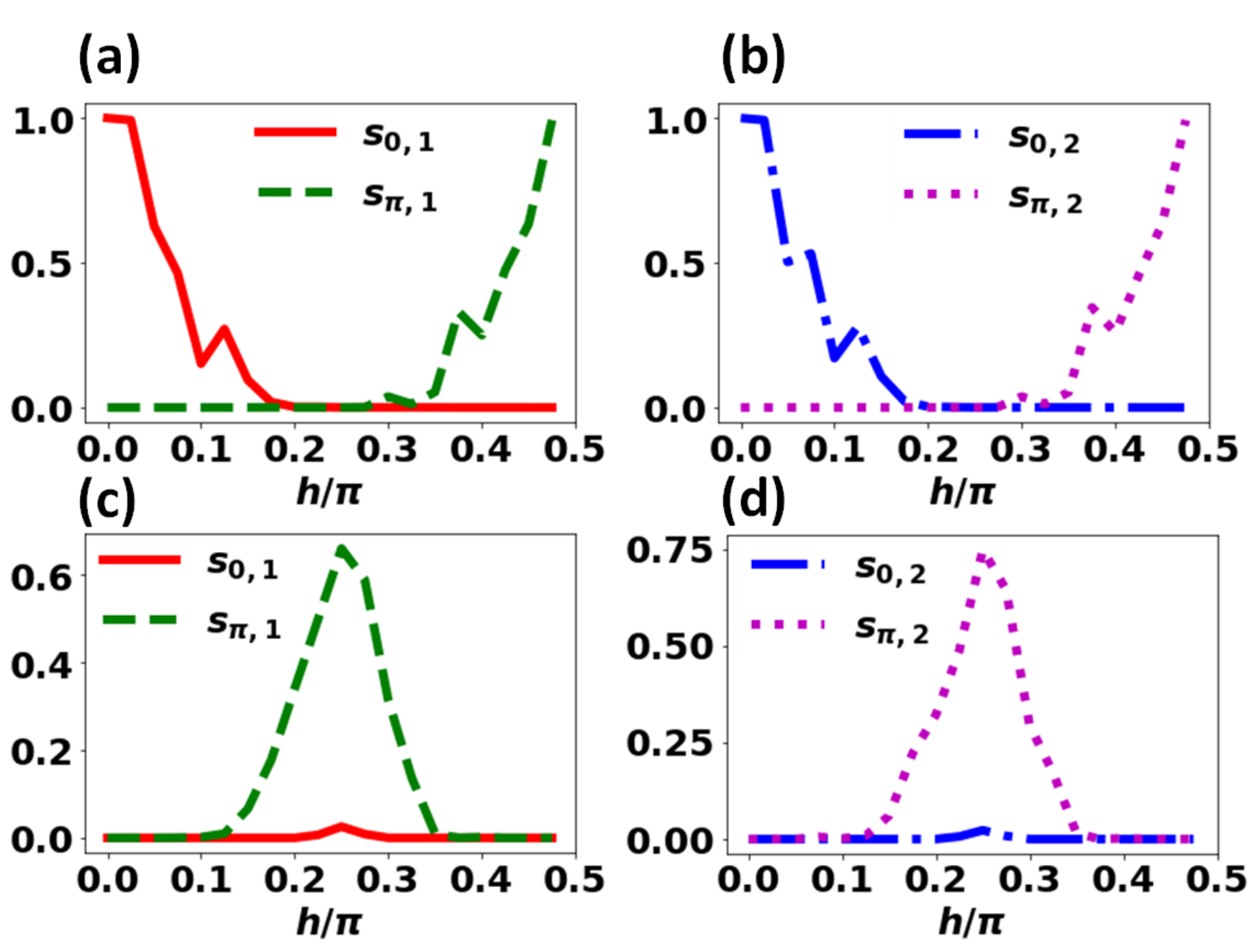}
		\caption{The four corner spectral functions (with $\chi=16$ and $\epsilon=0.01$) plotted as a function of (we set $T=2$) (a,c) $h$ and (b,d) $J\equiv J_y$. The system size is taken as $4\times 2$ and the other systems parameters are $J_x=0.05\pi/2$, (a,c) $J_y=0.3\pi$ and (b,d) $h=0.4\pi$.}
		\label{spectral}
	\end{figure}
\end{center}

Since $\gamma_{A,1,1}$ and $\gamma_{B,N_x,N_y}$ represent exact $\pi/T$ quasienergy excitations in the ideal $h=\pi/2$ case, any $|\varepsilon_m\rangle $ is mapped onto some $|\varepsilon_n\rangle $ with $|\varepsilon_n-\varepsilon_m|=\pi/T$. It then follows that $s_{0,1}=s_{0,2}=0$, \JB{but $s_{\pi,1}=s_{\pi,2}=1$}. Away from the ideal case, $\gamma_{A,1,1}$ and $\gamma_{B,N_x,N_y}$ no longer represent the system's true corner MPMs. However, if corner MPMs still exist, they are expected to have large support on $\gamma_{A,1,1}$ and $\gamma_{B,N_x,N_y}$, thus yielding $s_{0,1},s_{0,2}\approx0$ and $s_{\pi,1},s_{\pi,2}\approx 1$. \R{To this end, since these corner MPMs, if exist, are not exactly $\pi/T$ quasienergy excitations at finite system sizes, a state $|\varepsilon_n\rangle$ may be mapped onto $|\varepsilon_m\rangle$ with  $|\varepsilon_n-\varepsilon_m|=\Delta \varepsilon_{n,m}\neq \pi/T$. This motivates the integration of $S_1(\varepsilon_n,\eta)$ and $S_2(\varepsilon_n,\eta)$ with respect to $\eta$ over a small quasienergy window centered around $\pm \pi/T$ in the definition of $s_{\pi,1}$ and $s_{\pi,2}$ above. Moreover, since $\Delta \varepsilon_{n,m}$ may not be the same over all Floquet eigenstates, the sampling of $S_1(\varepsilon_n,\eta)$ and $S_2(\varepsilon_n,\eta)$ over several Floquet eigenstates is necessary. This justifies the summation over $\chi$ appearing in $s_{\pi,1}$ and $s_{\pi,2}$ above. Similar consideration is taken into account in defining $s_{0,1}$ and $s_{0,2}$ to properly probe the presence of zero quasienergy excitations.} We numerically evaluate these corner spectral functions in Fig.~\ref{spectral} for a system of size $4\times 2$ under periodic boundary conditions (PBC) in both directions. In particular, not only do we verify the robustness of the corner MPMs over a considerable range of parameter values, we also find evidence that the range of parameter values over which these corner MPMs exist matches the phase diagram of the system's 1D counterpart, i.e., Fig.~\ref{result1}(a). That is, by fixing $J_y=0.6\pi/T$ in panel (a), we observe that $s_{\pi,1}$ and $s_{\pi,2}$ will only become close to zero for $h<0.6\pi/T\ (T=2)$. Again consistent with the phase diagram in Fig.~\ref{result1}(a), for $h<0.4\pi/T\ (T=2)$, the values of $s_{0,1}$ and $s_{0,2}$ shoot up, thus hinting the existence of corner MZMs.  Likewise, by fixing $h=0.8\pi/T$ in panel (b), we observe that $s_{\pi,1}$ and $s_{\pi,2}$ remain finite for $0.2\pi/T <J_y < 0.8\pi/T\ (T=2)$.  \JB{These detailed analysis hence clearly indicates that the MPMs (as seen below, MZMs are not of much interest in the present study)  in the original one-dimensional spin chain survives in the spin ladder system, but now in the form of second-order topological corner modes over a considerable range of system parameters.} 

It is then interesting to connect the above-identified corner MPMs with the subharmonic response of the system, specifically, the $2T$ magnetization dynamics. This angle of analysis is certainly motivated by how DTCs are examined.  Further, in a recent experiment on an acoustic platform \cite{Zhangexp}, it is observed that a topological corner $\pi$ mode can be used to generate stable period-doubling oscillations after a corner excitation.  Certainly, the subharmonic response in our spin ladder system is of many-body nature; it involves the bulk many-body states in general because the MPMs are highly nonlocal in the spin representation.
To present our computational findings regarding  the 2T magnetization dynamics,  we first define the system's magnetization as
\begin{equation}
\langle S_z (t)\rangle = \sum_{i=1}^{N_x} \sum_{j=1}^{N_y} \langle \sigma_{z,i,j} \rangle \;. \label{magnetization}
\end{equation}

\subsection{Effect of the system size on the sub-harmonic response in 1D}

\JB{The subharmonic response associated with genuine DTCs is known to exhibit an infinite life-time in the thermodynamic limit.  With this insight it would be interesting to discuss how the $2T$ magnetization dynamics in our clean spin ladders changes upon increasing the system size.
To that end we start with the 1D limit} of Eq.~(\ref{sys}), i.e., by taking $N_x=1$, and analyze the effect of increasing the system size $N_y$. First, note that MPMs generally have a finite correlation length. This results in some overlap between two MPMs localized at the opposite ends, which prevents them from being true $\pi/T$ quasienergy excitations. This explains the imperfect $2T$ magnetization dynamics with a finite lifetime observed at $h\neq \pi/2$ and small system size (see e.g. Fig.~\ref{result1}(b)). Increasing the system size reduces the overlap between two MPMs, which in turn brings them closer to being $\pi/T$ quasienergy excitations and leads to improvement in the $2T$ magnetization dynamics and its lifetime up to a critical system size $N_{y,\rm crit}$ (see Fig.~\ref{result1}(c)). 

Remarkably, increasing the system size beyond $N_{y,\rm crit}$ is found to weaken the $2T$ magnetization dynamics.  This finding may be partially understood as follows. At $hT/2=\pi/2$, the initial state $|\uparrow\cdots \uparrow \rangle $ can be exactly written as a superposition of two Floquet eigenstates with quasienergies $-N_y J_y$ and $-N_y J_y+\pi/T$. In particular, at sufficiently small system sizes, these two quasienergies are generally nondegenerate. However, due to the periodicity of the quasienergy Brillouin zone $\varepsilon \in \left( -\pi/T, \pi/T\right]$, the quasienergies $-N_y J_y$ and $-N_y J_y+\pi/T$ are only defined modulo $2\pi/T$. As the system size ($N_y$) becomes sufficiently large, these values may loop back in the quasienergy Brillouin zone. This, coupled with the fact that the number of quasienergy eigenstates scales up significantly with the system size,  strengthens the chance of degeneracy or near-degeneracy between many pairs of Floquet eigenstates. If such type of unavoidable degeneracy involves states differing by a single spin excitation, it becomes extremely sensitive to perturbation. Indeed, a small deviation from $hT/2=\pi/2$ may result in considerable hybridization between two such Floquet eigenstates, thus splitting their quasienergy by an arbitrary value. This in turn results in the destruction of $\pi/T$ quasienergy pairing and, consequently, $2T$ magnetization dynamics (see Fig.~\ref{result1}(d)).     

\JB{One may also appreciate the loss of \MS{period doubling dynamics} in terms of  the movement of the domain wall along the spin chain.  At $hT/2=\pi/2$, as long as the state has non-vanishing magnetization, e.g. the product state in the $\sigma_z$ basis $|\psi_0\rangle=|\downarrow\uparrow\cdots\uparrow\uparrow\rangle$, then the period doubling dynamics is self-evident. Indeed, in this case the Floquet operator $U$ can rewritten as $U=P_xU_p,$ where $P_x$ represents a perfect spin flip operator $$P_x=e^{-i\pi/2\sum_j\sigma_{x,j}}$$ and the dynamics emanating from one eigenstate of $U_p$ with nonvanishing magnetization apparently displays $2T$ oscillation dynamics.   For the same state but with $h=(\pi/T+\delta h)T/2\neq \pi/2$, we may first consider the zero-th order approximation to the effective Floquet Hamiltonian of $U_p=e^{-iH_{p,F}T}$, namely,
 $$H_{p,F}^{(0)}= -J_y/2\sum_j \sigma_{z,1,j} \sigma_{z,1,j+1} - \delta h/2\sum_j\sigma_{x,j},$$ which is the transverse field Ising Hamiltonian (TFIH). For  $\delta h<J_y$, the TFIH is in the ferromagnetic phase, where we have an almost degenerate ground state separated by a finite energy gap from all other excited states.  With the presence of the perfect spin flip operator $P_x$ in $U$, the originally degenerate states in $H_{p,F}^{(0)}$ take us to other Floquet states of $U$ with a quasienergy separation of $\pi/T$ in $U$.   The $\delta h$ term however induces the movement of the domain wall, initially introduced in the state $|\psi_0\rangle$,  along the spin chain.  Indeed, using again Jordan-Wigner transformation and Bogoliubov transformation, the $\delta h$ term behaves likes a hopping term for a domain wall. 
 Thus, tuning away $h$ from the peculiar value $\pi/T$ generates a mobile domain wall, without a direct energy cost \MS{to the first order approximation}.  This indicates that $|\psi_0\rangle$ is no longer an eigenstate of $U_p$, and
 $|\psi_0\rangle$ under the action of $U_p$ alone can hence diffuse into a linear combination of different states with the domain wall at different locations. Therefore, the joint action of $P_x$ and $U_p$ cannot bring the state back after two periods, hence equivalently there is loss of $\pi/T$ pairing. \MS{In addition, these delocalized domain walls fluctuate and thus destroy the long-range order in the state, leading to a vanishing value of magnetization. Therefore, only the state with spins all up or all down in the 1D model can display the period doubling dynamics for a suffiecintly long time. However, with hybridization between different Floquet states, i.e. Floquet resonance, as discussed previously, domain walls inevitably emerge in the state under Floquet driving.}}

The above-analyzed loss of \MS{persistent period doubling dynamics} can generally be avoided in the presence of strong disorder \cite{MBL1,MBL2,MBL3,MBL4} which \MS{leads to localization of the domain walls and prevents Floquet heating}, or nonlocal interaction \cite{DTC9}. Together with the robust $\pi/T$ quasienergy pairing due to Majorana $\pi$ modes and their topological protection, one may then obtain genuine DTC phases with disorder.  Though the absence of any of these mechanisms precludes the clean systems under consideration here from being a true DTC, a better understanding of the subharmonic responses in a clean setting is still of great interest, especially to the quantum simulation of clean periodically driven systems.  

\section{$2T$-magnetization dynamics of a driven spin ladder}
 \JB{In this section, we shall show that a periodically driven spin ladder allows for an inherent mechanism for avoiding the above-mentioned dangerous degeneracy up to moderate system sizes. This new insight is interesting, because it suggests that the dynamics of driven spin ladders resembles that of DTCs at larger system sizes than in  the corresponding 1D spin chains. }	 

\subsection{Comparison with the 1D model}
To establish a simple and fair comparison between the spin ladder dynamics and the associated 1D model, we consider the model of Eq.~(\ref{sys}) at $N_x,N_y=2,2$ and $N_x,N_y=1,4$ respectively. 

\begin{center} 
	\begin{figure}
		\includegraphics[scale=0.5]{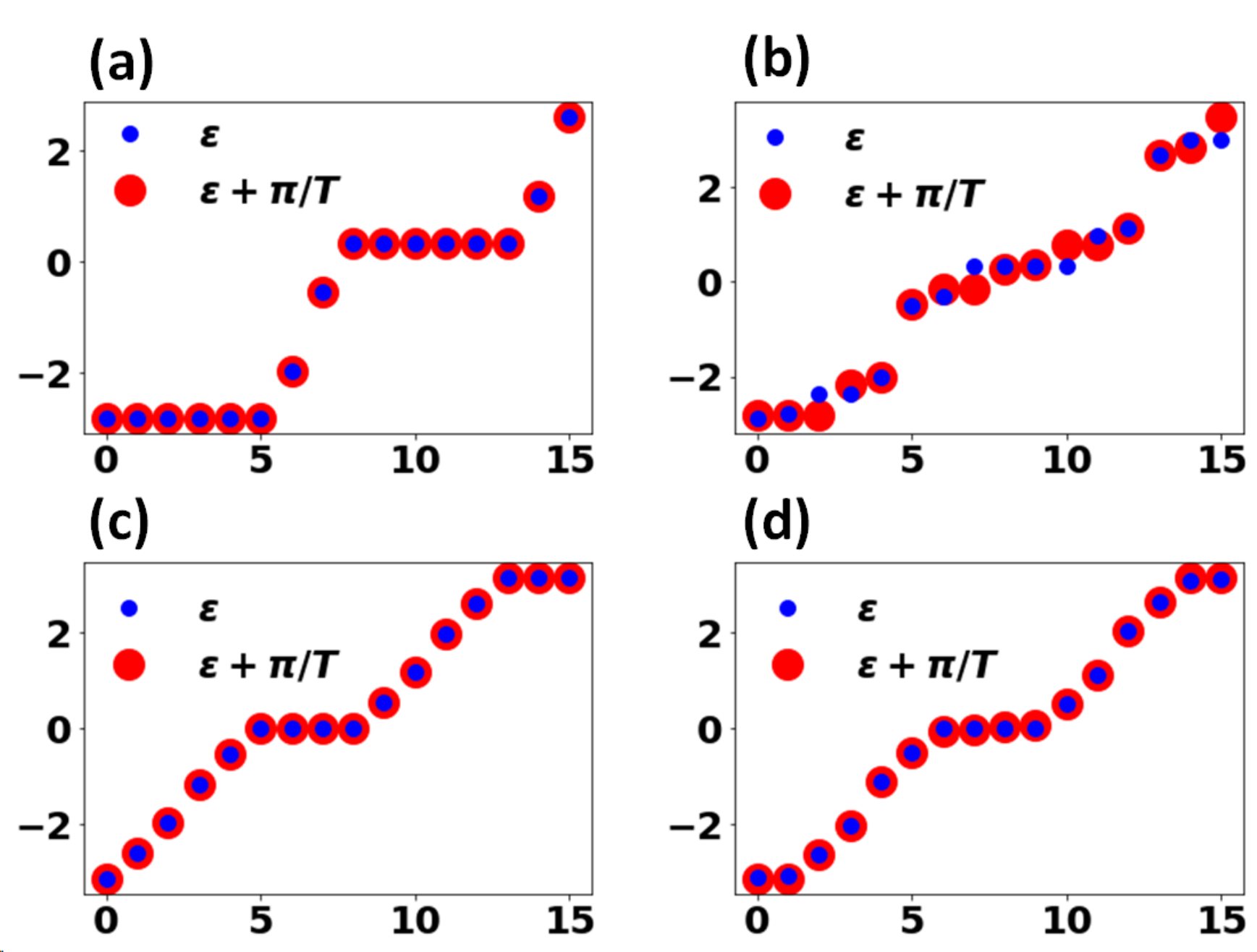}
		\caption{Quasienergy levels (and their $\pi/T$-shifted values) associated with system configurations (a,b) $N_x, N_y=1, 4$ and (c,d) $N_x,N_y=2,2$. We take (a,c) a solvable parameter value of $h=\pi/T$ associated with perfect $2T$ magnetization dynamics and (b,d) a perturbed value of $h=0.85 \pi/T$.} 
		\label{qspec}
	\end{figure}
\end{center}

We start by considering the system configuration of $N_x,N_y=1,4$. At $hT/2=\pi/2$, the system's $16$ quasienergies can be explicitly obtained as $\varepsilon_{s_1,s_2,s_3,+}= -J_y \sum_{j=1} s_j$ and $\varepsilon_{s_1,s_2,s_3,-}= -J_y \sum_{j=1} s_j +\pi/T$, where $s_1,s_2,s_3=\pm 1$. At sufficiently small $J_y$, the quasienergies $\varepsilon_{1,1,1,\pm}$, which are associated with eigenstates $|\varepsilon_{1,1,1,\pm}\rangle \propto |\uparrow \uparrow \uparrow \uparrow \rangle \pm |\downarrow \downarrow \downarrow \downarrow \rangle$, may be nondegenerate. On the other hand, the other Floquet eigenstates are at least twofold degenerate, e.g., $\varepsilon_{+,-,-,+} = \varepsilon_{-,+,-,+}$ (see Fig.~\ref{qspec}(a, b) for the complete quasienergy spectrum). In particular, these degenerate Floquet eigenstates involve states differing by a single spin excitation, e.g., $|\uparrow \downarrow \uparrow \uparrow \rangle$ and $|\uparrow \downarrow \downarrow \uparrow \rangle$, which can be easily split by perturbation (see Fig.~\ref{qspec}(b)). As an immediate consequence numerically confirmed below, while choosing $|\uparrow\cdots \uparrow \rangle$ or $|\downarrow\cdots \downarrow \rangle$ as an initial state yields a robust $2T$ magnetization dynamics with respect to variation in the parameter $h$ (see Fig.~\ref{result1}(b)), other choices of initial states, which have significant support on the above degenerate Floquet eigenstates, no longer display the same robustness.

We now turn our attention to the system configuration of $N_x,N_y=2,2$. At $hT/2=\pi/2$, such a system is also analytically solvable with quasienergies $\varepsilon_{s_1,s_2,s_3,+}= -J_y (s_1+s_2)- J_y'(1+s_1s_2)s_3$. Similarly to the $N_x, N_y=1,4$ setting, Floquet eigenstates degeneracy exist except for $|\varepsilon_{-1,-1,-1,\pm}\rangle = |\uparrow \downarrow \downarrow \uparrow \rangle \pm |\downarrow \uparrow \uparrow \downarrow \rangle$ and $|\varepsilon_{1,1,1,\pm}\rangle = |\uparrow \uparrow \uparrow \uparrow \rangle \pm |\downarrow \downarrow \downarrow \downarrow \rangle$. However, in the $N_x,N_y=2,2$ configuration, the quasienergy degeneracy only involves states differing by at least two spins excitations. Since the source of perturbation in the present model comes from variation in the $h$ parameter characterizing a single spin rotation in the $x$-direction, such a degeneracy cannot be easily lifted, e.g., compare Fig.~\ref{qspec}(c) with Fig.~\ref{qspec}(d). 

In Table~\ref{T1}, we compare the minimum and maximum quasienergy spacing (relative to $\pi/T$) at various system sizes. Consistent with the above analysis, the spin ladder configurations outperform their 1D counterpart in terms of having quasienergy spacing closer to $\pi/T$ at moderate imperfection $h\neq \pi/T$. Increasing the system size generally brings the quasienergy spacing closer to $\pi/T$. Exceptions to this pattern can be attributed to the presence of accidental degeneracy between states differing by a single spin excitation, which becomes more probable at larger system sizes due to the periodicity of the quasienergy Brillouin zone and the larger number of Floquet eigenstates. 

\begin{center}
\begin{table}[]
    \centering
    \begin{tabular}{|c|c|c||c|c|c|}
    \hline
    Size & Min & Max & Max & Min & Size \\ \hline
        $2 \times 2$ & $1.20\times 10^{-3}$ & $0.023$ & $0.15$ & $1.81\times 10^{-3}$ & $1\times 4$\\
        $3\times 2$ & $5.42\times 10^{-6}$ & $0.016$ & $0.086$ & $2.08\times 10^{-4}$ & $1\times 6$\\
        $4\times 2$ & $2.06\times 10^{-5}$ & $0.017$ & $0.050$ & $2.49\times 10^{-5}$ & $1\times 8$ \\
        $5\times 2$ & $2.14\times 10^{-6}$ & $0.013$ & $0.032$ & $1.38\times 10^{-4}$ & $1\times 10$ \\
        $6\times 2$ & $4.53\times 10^{-7}$ & $0.015$ & $0.017$ & $1.37\times 10^{-5}$ & $1\times 12$ \\ \hline
    \end{tabular}
    \caption{Maximum and minimum deviation of the quasienergy spacing from $\pi/T$ at various system sizes. The system parameters are chosen as $h=0.85\pi/T$, $J_x=0.05\pi/T$, and $J_y=2/T$.}
    \label{T1}
\end{table}
\end{center}

\subsection{Computational results}
\begin{center} 
	\begin{figure}
		\includegraphics[scale=0.5]{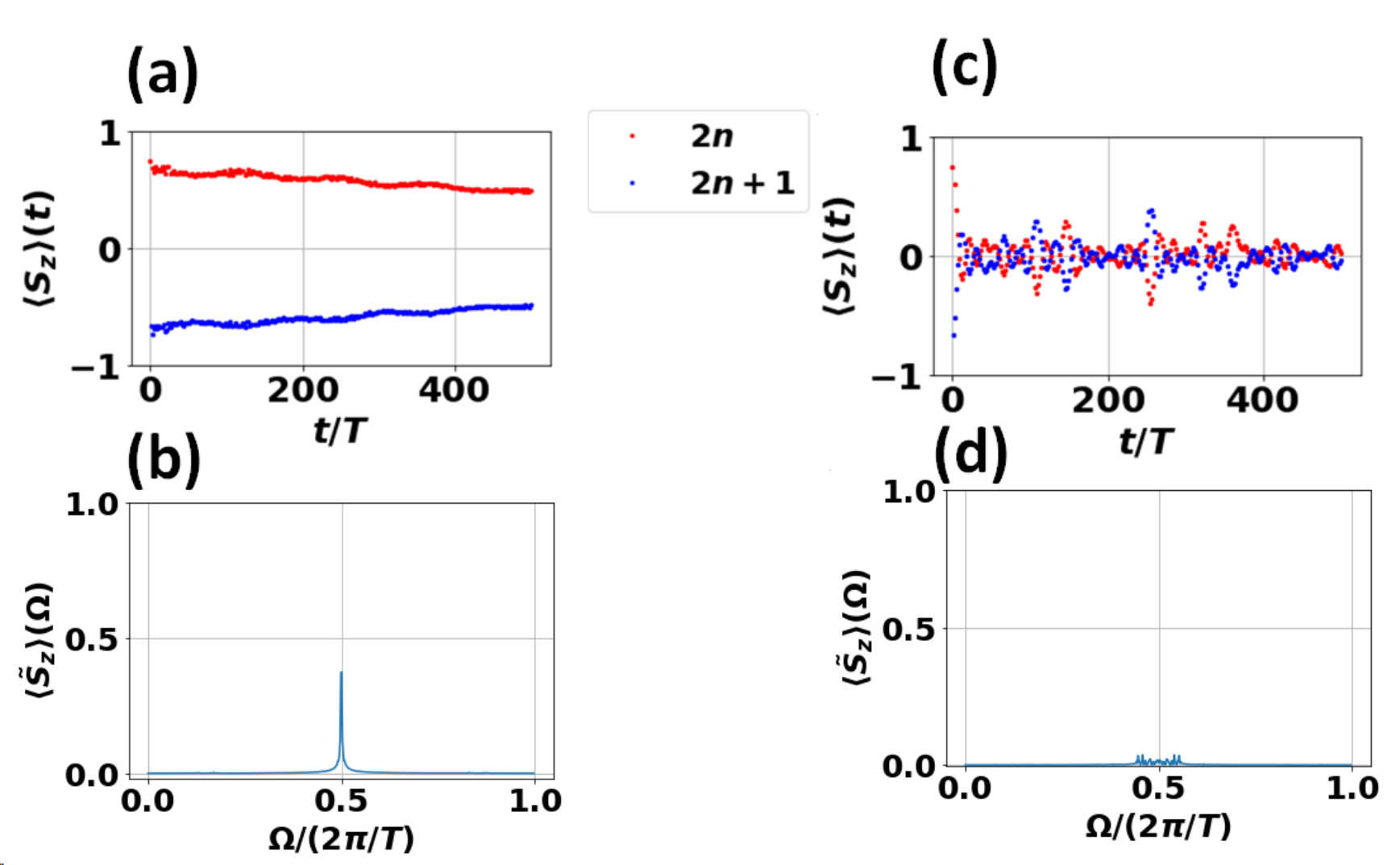}
		\caption{The stroboscopic magnetization dynamics (a, c) and its associated power spectrum (b, d) under the initial state of $|\uparrow \downarrow \uparrow \cdots \uparrow\rangle$. Panel (a, b) and (c, d) correspond to the configurations $N_x, N_y= 4\times 2$ and $1\times 8$ respectively. All parameter values are taken as $J_x=0.05\pi/T$, $h=0.85\pi/T$, and $J_y=2/T$.}
		\label{result5}
	\end{figure}
\end{center}

To support the above argument, we plot in Fig.~\ref{result5} the stroboscopic magnetization dynamics (see Eq.~(\ref{magnetization}))and its associated power spectrum, i.e., $\langle \tilde{S}_z\rangle = \frac{1}{M}\sum_{n=1}^M \langle S_z (nT) \rangle e^{-\mathrm{i} \Omega n T/M}$, at configurations $1\times 8$ and $4\times 2$ under the initial state of $|\uparrow \downarrow \uparrow \cdots \uparrow\rangle$. Indeed, while $2T$-periodicity remains robust in the $4\times 2$ setting, it is no longer visible in the purely 1D system.

\begin{center} 
	\begin{figure}
		\includegraphics[scale=0.5]{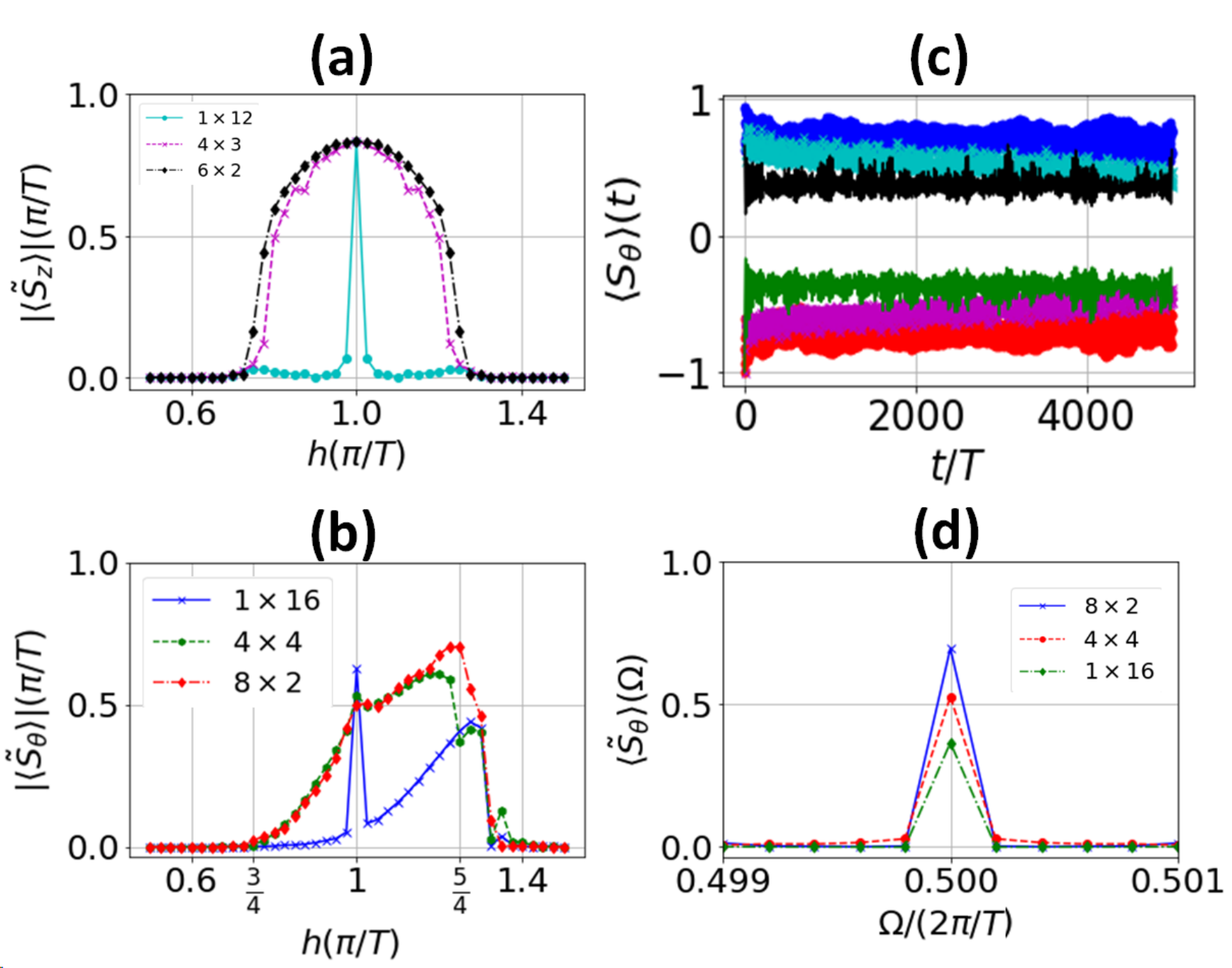}
		\caption{(a,b) The subharmonic response of the system's average magnetization at $\Omega=\pi/T$ and various lattice configurations under two different choices of initial states (a) $|\uparrow \downarrow \uparrow \cdots \uparrow \rangle $ and (b) $|\nearrow \cdots \nearrow \rangle$. (c,d) The stroboscopic evolution of the average magnetization in the $\hat{n}_{\pi/8}$ direction and its associated power spectrum for the initial state $|\nearrow \cdots \nearrow \rangle$. There, green/black, magenta/cyan, and red/blue respectively correspond to system sizes $1\times 16$, $4\times 4$, and $8\times 2$. In all panels, the other systems are chosen as $J_x=0.05\pi/T$, $J_y=0.6\pi/T$, and (c,d) $h=1.225\pi/T$.}
		\label{result3}
	\end{figure}
\end{center}

In Fig.~\ref{result3}(a, b), we further demonstrate the robustness of the $2T$ magnetization dynamics in the proposed $N_x\times N_y$ system by scanning its $\pi/T$ subharmonic peak over the parameter $h$ at two different initial states. First, we consider an initial state in which all but one spins is in the up ($+\hat{z}$) position, with the subharmonic response of the system's averaged magnetization shown in Fig.~\ref{result3}(a).  In this case, even a considerably large 1D lattice no longer demonstrates robust subharmonic response, as evidenced by the very narrow window of subharmonic peaks with respect to variation in a system parameter observed in Fig.~\ref{result3}(a). By contrast, in the spin ladder configurations, a significantly large subharmonic response is still observed over a much broader range of the parameter $h$.

Second, we consider an initial state in which all spins are aligned in a diagonal direction, e.g., $+\hat{n}_{\pi/8}=\cos(\pi/4)\hat{z}+\sin(\pi/4)\hat{y}$, then stroboscopically evaluate the magnetization in the same direction and present our results in Fig.~\ref{result3}(b). Here, we find that $h=\pi/T$ generically no longer represents the value at which the subharmonic response is maximum.  The relatively large subharmonic response of 1D configuration $1\times 16$ at $h=\pi/T$ is accidental, because the associated magnetization profile actually exhibits periodicity larger than $2T$ (See Appendix~\ref{app2}). Such a $>2T$ periodicity is merely a result of some fine tuning of parameter values and quickly disappears as $h$ is tuned away from $\pi/T$. As all spins are aligned at an angle of $\pi/4$ from the $+\hat{z}$-axis, the genuine maximum $2T$ subharmonic response is instead observed around $h=1.25\pi/T$, which can intuitively be understood as follows. Ignoring the interaction terms, a Floquet operator at $h=5\pi/(4T)$ effectively rotates the system state towards $|\uparrow\cdots \uparrow \rangle$, which is known to have considerable support on two nondegenerate Floquet eigenstates with quasienergy separation of $\pi/T$.  In the vicinity of $h=1.25\pi/T$, the magnetization profile vs time and the associated power spectrum are also shown in Fig.~\ref{result3}(c) and (d).    

\section{Concluding remarks}

\JB{We have investigated the topological and dynamical features of  periodically driven spin ladder configurations up to moderate system sizes.  Aligned with how the dynamics of genuine DTCs is analyzed,  we examine the robustness and time scale of subharmonic response to the periodic driving,  in connection with the presence of MPMs at two corners in the equivalent Majorana lattice models. While such MPMs may become extremely nonlocal in the original spin representation, their role as quasienergy $\pi/T$ excitations ensures the system's full Floquet eigenstates to form $\pi/T$ quasienergy spacing, which then leads to $2T$ magnetization dynamics over a finite but considerable time scale.  The topological nature of such corner MPMs leads to the robustness of the obtained $2T$ periodic feature over a range of parameter values. The subharmonic response of spin ladders can survive at larger system sizes than its 1D counterpart, with less sensitivity to the choice of initial states.  We have also highlighted the role of spin-spin interaction in the rung direction in removing the degeneracy between states differing by a single spin excitation.  }

Due to the simplicity of our proposed model and its similarity with existing 1D DTC models, its experimental realization is straightforward. In particular, there exists various platforms on which periodically driven spin systems are already experimentally realized, which include trapped ions \cite{DTCexp1,DTCexp7}, Nitrogen-Vacancy centers in diamonds \cite{DTCexp2}, and nuclear magnetic moments \cite{DTCexp3,DTCexp4,DTCexp5}. Among these available platforms, trapped ions of Ref.~\cite{DTCexp1,DTCexp7} are the most suitable platform due to the flexibility in arranging the $^{171}Yb^+$ ions into various different lattice configurations, whereas their experimental execution may simply follow directly the method presented in Ref.~\cite{DTCexp1,DTCexp7}.  Alternatively, the emerging Sycamore processor \cite{DTCqs2,Sycamore} represents another promising platform to simulate various properties of DTCs or their related models. In particular, the ZZ interaction required to implement Eq.~(\ref{sys}) can be implemented using the native gates within the Sycamore processor \cite{DTCqs2}.   

This work opens up interesting avenues for future work. First, following the intricate interplay between MPMs and subharmonic response unveiled in this work, it is of interest to investigate subharmonic response in a class of spin systems associated with a type of periodically driven topological superconductor, whose number of MPMs is tunable by some system parameters \cite{FMF6,FMF7,FMF8}. Second, exploring the effect of increasing the number of connectivities for each spin is another direction worth pursuing. In particular, with each spin possessing either three or four connectivities (depending on the boundary conditions) in the model considered in this paper, accidental degeneracy involving states differing by two spin excitations may still remain. It is expected that a modification to Eq.~(\ref{sys}) by adding more spin connectivities will further remove this less dangerous degeneracy and hence may have significant impact on the dynamical features.  Finally, our main construction of disorder-free system above can be naturally extended to simulating other types of subharmonic response characterized by $nT$-periodic observables \cite{repDTC}, where $n\in \mathbb{Z}$.

\vspace{0.3cm}
	
	\begin{acknowledgements}
		{\bf Acknowledgement}: R.W.B. is supported by the Australian Research Council Centre of Excellence for Engineered Quantum Systems (EQUS, CE170100009).  R.W.B. thanks Isaac Kim and Adrian Chapman for useful discussions.  J.G. acknowledges fund support by the Singapore NRF Grant No. NRFNRFI2017-04 (WBS No. R-144-000-378-281).
	\end{acknowledgements}

\appendix 

\section{MPMs in periodically driven spin-1/2 chain under OBC and PBC} \label{app1}

We define a set of Majorana operators via a generalized JW transformation 
\begin{eqnarray}
\gamma_{A,i,j}&=&\prod_{(l<i),(m\leq N_y)} \sigma_{x,l,m} \prod_{n<j} \sigma_{x,i,n} \sigma_{z,i,j} \;, \nonumber \\
\gamma_{B,i,j}&=& \prod_{(l<i),(m\leq N_y)} \sigma_{x,l,m} \prod_{n<j} \sigma_{x,i,n} \sigma_{y,i,j} \;, \label{JW} 
\end{eqnarray}
which satisfy $\left\lbrace\gamma_{S,i,j},\gamma_{S',i',j'}\right\rbrace= 2\delta_{S,S'}\delta_{i,i'}\delta_{j,j'}$. In terms of Majorana operators $\gamma_{A,i,j}$ and $\gamma_{B,i,j}$, the 1D limit ($N_x=1$) of the Floquet operator of Eq.~(3) in the main text can be written as
\begin{eqnarray}
U_{\rm 1D}^{\rm (OBC)} &=& e^{-\sum_{i=1}^N  h\gamma_{A,j}\gamma_{B,j}} e^{\sum_{i=1}^N  J_x \gamma_{B,j}\gamma_{A,j+1}}\nonumber \\
U_{\rm 1D}^{\rm (PBC)} &=&U_{\rm 1D}^{(OBC)} \times e^{ J_x \gamma_{B,1}\left(\prod_{1<j<N} (-\mathrm{i})\gamma_{A,j}\gamma_{B,j}\right)\gamma_{A,N}} \;, \nonumber \\
\end{eqnarray}
under OBC and PBC respectively.

\subsection{OBC case}

Let us first analyze the existence of MPMs in the OBC setting. Without loss of generality, we shall only look for an MPM localized near the left end. To this end, We first note that (taking $T=2$ throughout)
\begin{eqnarray}
U_{\rm 1D}^{\rm (OBC)\dagger} \gamma_{A,1} U_{\rm 1D}^{\rm (OBC)} &=& \cos(2h) \gamma_{A,1} -\sin(2h) \gamma_{B,1} \;, \nonumber \\
U_{\rm 1D}^{\rm (OBC)\dagger} \gamma_{B,N} U_{\rm 1D}^{\rm (OBC)} &=& \cos(2h) \gamma_{B,N}+ \sin(2h) \gamma_{A,N} \;, \nonumber \\
U_{\rm 1D}^{\rm (OBC)\dagger} \gamma_{A,j>1} U_{\rm 1D}^{\rm (OBC)} &=& \cos(2J_x)\left[\cos(2h) \gamma_{A,j} \right. \nonumber \\
&-& \left. \sin(2h) \gamma_{B,j}\right] \nonumber \\
&+&\sin(2J_x)\left[\cos(2h) \gamma_{B,j-1} \right. \nonumber \\
&+& \left. \sin(2h) \gamma_{A,j-1}\right] \;, \nonumber \\
U_{\rm 1D}^{\rm (OBC)\dagger} \gamma_{B,j<N} U_{\rm 1D}^{\rm (OBC)} &=& \cos(2J_x)\left[\cos(2h) \gamma_{B,j} \right. \nonumber \\
&+& \left. \sin(2h) \gamma_{A,j}\right] \nonumber \\
&-&\sin(2J_x)\left[\cos(2h) \gamma_{A,j+1} \right. \nonumber \\
&-& \left. \sin(2h) \gamma_{B,j+1}\right] \;. \nonumber \\ \label{act} 
\end{eqnarray}	
An MPM, if it exists, can be written in the form (ignoring normalization)
\begin{equation}
\gamma_\pi =\gamma_{A,1} + \sum_{j>0} \left( a_j\gamma_{A,j+1} + b_j \gamma_{B,j}\right) \label{ansatz}
\end{equation}
and satisfies $U_{\rm 1D}^{\rm (OBC)\dagger} \gamma_\pi U_{\rm 1D}^{\rm (OBC)} = -\gamma_\pi$. By imposing the latter condition term by term, we obtain
\begin{widetext}
\begin{eqnarray}
a_1 &=& \frac{-1-\cos(2h)-\sin(2h)\cos(2J_x) b_1}{\sin(2h) \sin(2J_x)} \;, \nonumber \\
a_1 &=& \frac{-b_1 + \sin(2h) -\cos(2J_x)\cos(2h)b_1}{\sin(2J_x)\cos(2h)} \;, \nonumber \\
a_{j>1} &=& \frac{-a_{j-1}-\cos(2h) \cos(2J_x) a_{j-1} -\sin(2h)\cos(2J_x) b_j +\sin(2J_x)\cos(2h)b_{j-1}}{\sin(2h) \sin(2J_x)}  \;, \nonumber \\
b_{j>1} &=& \frac{-b_j-\cos(2h) \sin(2J_x) a_{j} +\sin(2h)\cos(2J_x) a_{j-1} -\sin(2J_x)\sin(2h)b_{j-1}}{\cos(2h)\cos(2J_x)}  \;, \label{MPMcond}
\end{eqnarray}
\end{widetext}
where the first and second lines are distinct equations obtained from the condition that $U_{\rm 1D}^{\rm (OBC)\dagger} \gamma_\pi U_{\rm 1D}^{\rm (OBC)}$ contains $-\gamma_{A,1}$ and $-b_1\gamma_{B,1}$ respectively. Equation~(\ref{MPMcond}) can be written in a matrix form as
\begin{equation}
\left(\begin{array}{c}
a_j \\
b_j
\end{array}\right) =M_{\rm OBC} \left(\begin{array}{c}
a_{j-1} \\
b_{j-1}
\end{array}\right) \;, \label{recur}
\end{equation}
where 
\begin{widetext} 
\begin{equation}
M_{\rm OBC} = \frac{1}{\sin(2h)\sin(2J_x)} \left(\begin{array}{cc}
-(1+2\cos(2h)\cos(2J)+\cos^2(2J_x)) & \sin(2J_x)\left[\cos(2h)+\cos(2J_x)\right] \\
\sin(2J_x)\left[\cos(2h)+\cos(2J_x)\right] & -\sin^2(2J_x) \\
\end{array}\right) \;.
\end{equation}
\end{widetext}
The eigenvalues of $M_{\rm OBC}$ are 
\begin{equation}
E_\pm^{(\rm OBC)} = -\frac{1+\cos(2h)\cos(2J)\mp \left[\cos(2h)+\cos(2J)\right]}{\sin(2h)\sin(2J)} \;, 
\end{equation}
with their corresponding eigenvectors being
\begin{equation}
\psi_+ = \left(\begin{array}{c}
\sin(J_x) \\
\cos(J_x)
\end{array}\right) \;, \; 
\psi_- = \left(\begin{array}{c}
\cos(J_x) \\
-\sin(J_x)
\end{array}\right) \;.
\end{equation}

By solving the first two lines of Eq.~(\ref{MPMcond}), we obtain
\begin{equation}
\left(\begin{array}{c}
a_1 \\
b_1
\end{array}\right)= -\frac{\cos(h)}{\sin(h)\sin(J_x)} \psi_- \;.
\end{equation}
Therefore, Eq.~(\ref{recur}) becomes
\begin{equation}
\left(\begin{array}{c}
a_j \\
b_j
\end{array}\right) =\left(E_-^{\rm (OBC)}\right)^{j-1} \left(\begin{array}{c}
a_1 \\
b_1
\end{array}\right) \;,
\end{equation}
and a well-defined (normalizable) MPM solution exists if $|E_-^{\rm(OBC)}|< 1$. In particular, it is easily verified that the line $h=\pi/2-J$ corresponds exactly to the $|E_\pm^{(\rm OBC)}|=1$, thus explaining the antidiagonal line separating the regime with and without MPMs in Fig.~2(a) of the main text. At $h>\pi/2-J$, $|E_-^{\rm(OBC)}|< 1$ and $|E_+^{\rm(OBC)}|> 1$, thus implying the presence of MPMs. At $h<\pi/2-J$, the opposite situation arises with $|E_-^{\rm(OBC)}|> 1$ and $|E_+^{\rm(OBC)}|< 1$, which therefore does not support MPMs. Finally, the full phase diagram of Fig.~2(a) in the main text can be reproduced by repeating the above analysis to show that MZMs (which satisfy $U_{\rm 1D}^{\rm (OBC)\dagger} \gamma_0 U_{\rm 1D}^{\rm (OBC)} = \gamma_0$) exist for $J>h$.

\subsection{PBC case}

We now turn our attention to the PBC setting. We first note that
\begin{widetext}
\begin{eqnarray}
U_{\rm 1D}^{\rm (PBC)\dagger} \gamma_{A,1} U_{\rm 1D}^{\rm (PBC)} &=& U_{\rm 1D}^{\rm (OBC)\dagger} \gamma_{A,1} U_{\rm 1D}^{\rm (OBC)} \;, \nonumber \\
U_{\rm 1D}^{\rm (PBC)\dagger} \gamma_{B,N} U_{\rm 1D}^{\rm (PBC)} &=& U_{\rm 1D}^{\rm (OBC)\dagger} \gamma_{B,N} U_{\rm 1D}^{\rm (OBC)} \;, \nonumber \\
U_{\rm 1D}^{\rm (PBC)\dagger} \gamma_{A,j>1} U_{\rm 1D}^{\rm (PBC)} &=&  \cos(2J_x) U_{\rm 1D}^{\rm (OBC)\dagger} \gamma_{A,j>1} U_{\rm 1D}^{\rm (OBC)} \nonumber \\
&+& \sin(2J_x) U_{\rm 1D}^{\rm (OBC)\dagger} \gamma_{\rm PBC} \gamma_{A,j>1} U_{\rm 1D}^{\rm (OBC)} \;, \nonumber \\
U_{\rm 1D}^{\rm (PBC)\dagger} \gamma_{B,j<N} U_{\rm 1D}^{\rm (PBC)} &=& \cos(2J_x) U_{\rm 1D}^{\rm (OBC)\dagger} \gamma_{B,j<N} U_{\rm 1D}^{\rm (OBC)}  \nonumber \\
&+&\sin(2J_x) U_{\rm 1D}^{\rm (OBC)\dagger} \gamma_{\rm PBC} \gamma_{B,j<N} U_{\rm 1D}^{\rm (OBC)} \;, \nonumber \\ \label{actpbc}
\end{eqnarray}
\end{widetext}
where $\gamma_{\rm PBC}=\gamma_{B,1}\left(\prod_{1<j<N} \gamma_{A,j}\gamma_{B,j}\right)\gamma_{A,N}$. In particular, since the Majorana lattice is now no longer a free fermion system, it is generally difficult to perform an extensive analytical treatment in the spirit of the OBC case above. We may however consider special points in the $0\pi$-PM and $\pi$-SG regime to understand how MPMs in these regimes are expected to be modified by the boundary ($\gamma_{\rm PBC}$) term. 

First, we note that along the $h=\pi/2$ line, which belongs to the $\pi$-SG regime, the two MPMs in the OBC case are precisely given by $\gamma_{A,1}$ and $\gamma_{B,N}$. As both of these MPMs commute with $\gamma_{\rm PBC}$, $U_{1D}^{PBC}$ and $U_{1D}^{OBC}$ act identically on $\gamma_{A,1}$ and $\gamma_{B,N}$. Consequently, they remain MPMs in the PBC case. It is thus expected that MPMs remain present throughout the $\pi$-SG regime in the PBC case.

Second, we note that along the $J_x=\pi/2$ line, which belongs to the $0\pi$-PM regime, the two MPMs in the OBC case are precisely given by $\left(\sin(h)\gamma_{A,1}+\cos(h)\gamma_{B,1}\right)$ and $\left(-\cos(h)\gamma_{A,N}+\sin(h)\gamma_{B,1}\right)$. Even in the PBC case, Eq.~(\ref{actpbc}) shows that weight-one Majorana operators are mapped onto other weight-one Majorana operators, rendering the Majorana representation of the system effectively free at this special $J_x$ value. Moreover, the set $\left\lbrace \gamma_{A,1}, \gamma_{B,1}\right\rbrace$ is mapped onto itself under conjugation by $U_{\rm 1D}^{\rm (PBC)}$. By assuming that a potential MPM takes the form of $\gamma_\pi = A \gamma_{A,1}+B\gamma_{B,1}$, the MPM condition $U_{1D}^{\rm (PBC)\dagger} \gamma_\pi U_{1D}^{\rm (PBC)}=-\gamma_\pi$ leads to the eigenvalue equation
\begin{equation}
\left(\begin{array}{cc}
\cos(2h) & -\sin(2h) \\
\sin(2h) & \cos(2h)
\end{array}\right) \left(\begin{array}{c}
A \\
B
\end{array}\right) = -\left(\begin{array}{c}
A \\
B
\end{array}\right) \;.
\end{equation}
However, the $2\times 2$ matrix on the left hand side only has eigenvalues of $\cos(2h)\pm \mathrm{i} \sin(2h)\neq -1$ for generic value of $h$. Consequently, MPMs do not exist along the $J_x=\pi/2$ line when the original spin system admits PBC. In this case, it is thus expected that MPMs are also absent throughout the whole $0\pi$-PM regime.

\section{Magnetization dynamics near $h=\pi/T$ under a $\pi/8$ rotated initial state} \label{app2}

\begin{center} 
	\begin{figure}
		\includegraphics[scale=0.5]{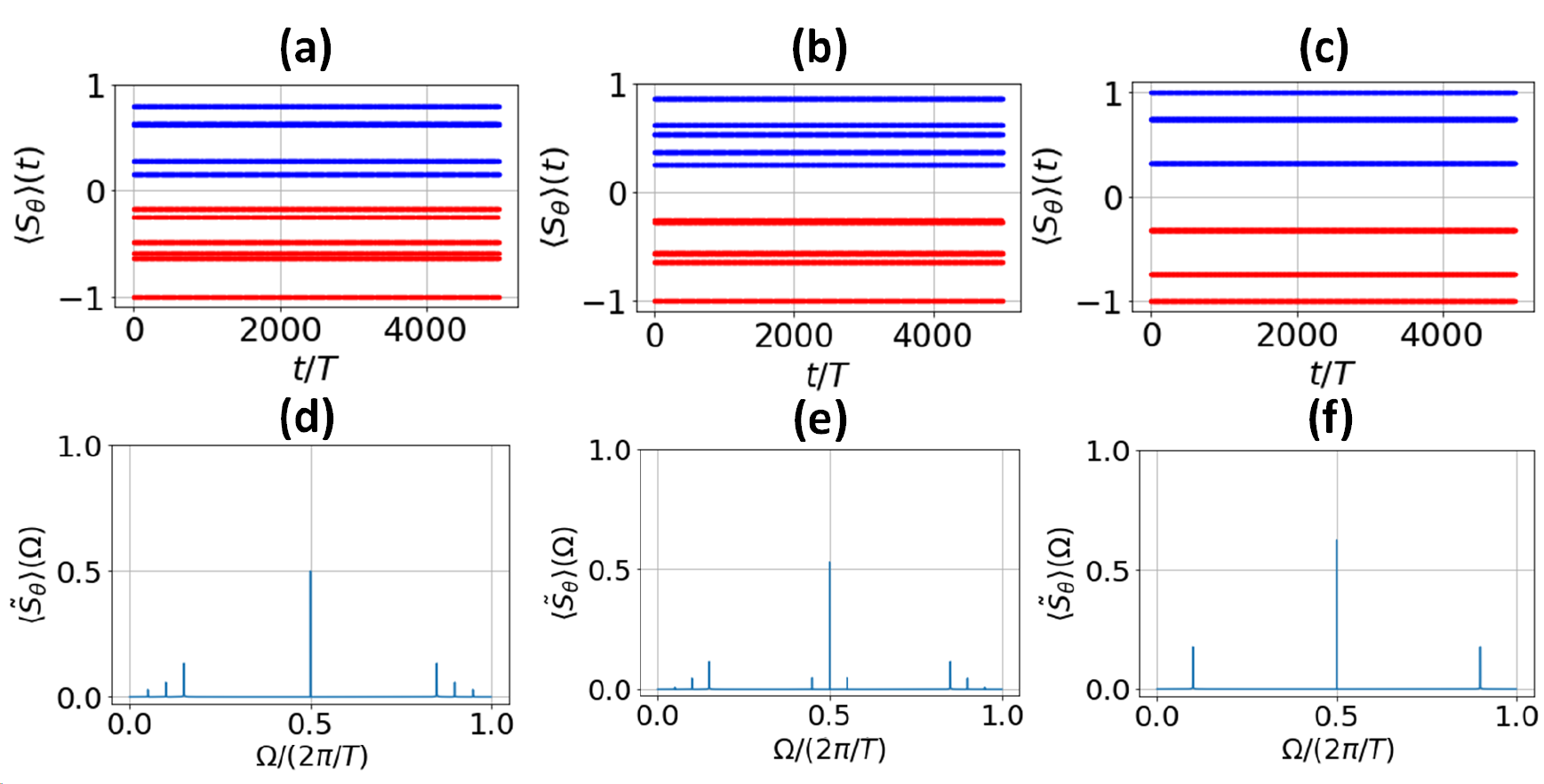}
		\caption{(a,b,c) The stroboscopic evolution of the system's average magnetization in the $\hat{n}_{\pi/8}$ direction for a spin ladder of size (a) $8\times 2$, (b) $4\times 4$, and (c) $1\times 16$. Blue (red) points are evaluated at odd (even) integer multiples of $T$, and the system is initially prepared in the $|\nearrow\cdots \nearrow \rangle$. (d,e,f) The associated power spectrum. System parameters are chosen as $J_x=0.05\pi/T$, $J_y=0.6\pi/T$, and $h=\pi/T$.}
		\label{rotated}
	\end{figure}
\end{center} 

\begin{center} 
	\begin{figure}
		\includegraphics[scale=0.5]{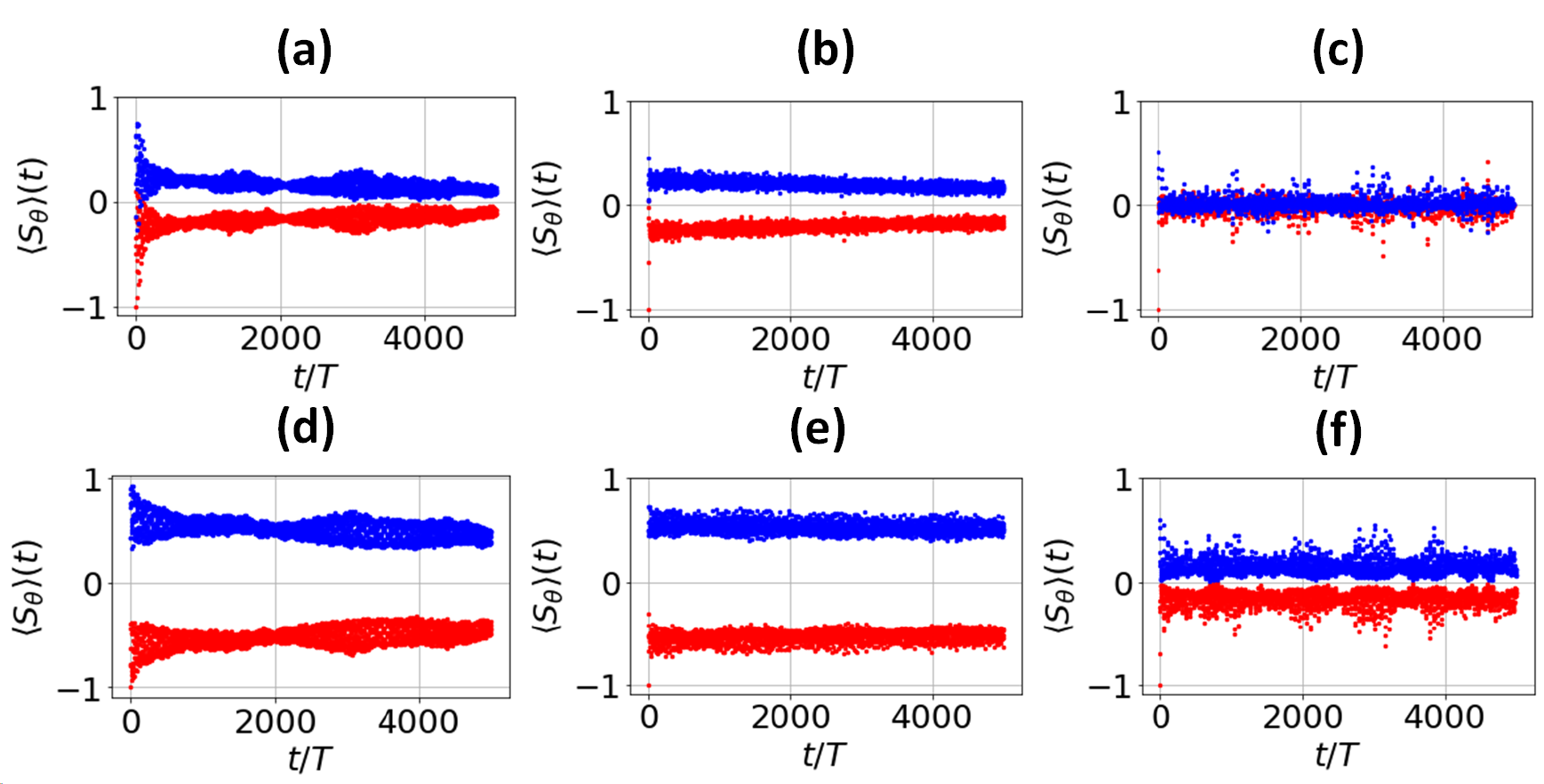}
		\caption{ The stroboscopic evolution of the system's average magnetization in the $\hat{n}_{\pi/8}$ direction for spin ladders of size (a,d) $8\times 2$, (b,e) $4\times 4$, and (c,f) $1\times 16$ at (a,b,c) $h=0.9\pi/T$ and (d,e,f) $h=1.1\pi/T$. Blue (red) points are evaluated at odd (even) integer multiples of $T$, and the system is initially prepared in the $|\nearrow\cdots \nearrow \rangle$. Other system parameters are the same as those in Fig.~\ref{rotated}.}
		\label{rotated2}
	\end{figure}
\end{center}

In Fig.~\ref{rotated}, we show the stroboscopic evolution of the system's average magnetization in the $\hat{n}_{\pi/8}$ direction at $h=\pi/T$ when all spins are initially aligned in the $\hat{n}_{\pi/8}$ direction. Indeed, as reported in the main text, despite the large $\pi/T$ subharmonic response of the system's average magnetization at $h=\pi/T$, its actual stroboscopic evolution actually exhibits different periodicity $>2T$ for the three lattice configurations under consideration. As evidenced in Fig.~\ref{rotated2}, such a $>2T$ periodicity quickly disappears and turns into the expected $2T$ periodicity as $h$ slightly deviates from the fine tuned value of $\pi/T$. In this case, the accidentally large $\pi/T$ subharmonic response at $h=\pi/T$ arises due to the symmetry of the magnetization spectrum oscillating between positive and negative values and simply represents an artefact of the power spectrum definition. Indeed, the larger peak observed for the configuration $1\times 16$ does not represent a sharper $2T$ periodicity as compared with the other two configurations, rather it only reflects the more symmetric structure of the observed magnetization profile about $\langle S_\theta \rangle=0$ due to its periodicity being an even integer multiple of $T$.   
	

\begin{thebibliography}{99}
		\bibitem{DTC1} K.~Sacha, \pra ~{\bf 91}, 033617 (2015).
		\bibitem{DTC2} D.~V.~Else, B.~Bauer, and C.~Nayak, \prl ~{\bf 117}, 090402 (2016).
		\bibitem{DTC3} D.~V.~Else, B.~Bauer, and C.~Nayak, Phys.~Rev.~X {\bf 7}, 011026
		(2017).
		\bibitem{DTC4} C.~W.~von~Keyserlingk and S.~L.~Sondhi, \prb ~{\bf 93},
		245146 (2016).
		\bibitem{DTC5}  V.~Khemani, A.~Lazarides, R.~Moessner, and S.~L.~Sondhi, \prl ~{\bf 116}, 250401 (2016).
		\bibitem{DTC6} N.~Y.~Yao, A.~C.~Potter, I.-D. Potirniche, and A.~Vishwanath, \prl ~{\bf 118}, 030401 (2017).
		\bibitem{DTC7} W.~W.~Ho, S.~Choi, M.~D.~Lukin, and D.~A.~Abanin, \prl ~{\bf 119}, 010602 (2017).
		\bibitem{DTC8} B.~Huang, Y.-H.~Wu, and W.~V.~Liu, \prl ~{\bf 120}, 110603 (2018).
		\bibitem{DTC9} A.~Russomanno, F.~lemini, M.~Dalmonte, and R.~Fazio, \prb ~{\bf 95}, 214307 (2017).
		\bibitem{DTC10} A.~Russomanno, B.~E.~Friedman, and E.~G.~D.~Torre, \prb ~{\bf 96}, 045422 (2017).
		\bibitem{DTC11} W.~C.~Yu, J.~Tangpanitanon, A.~W.~Glaetzle, D.~Jaksch, and
		D.~G.~Angelakis, \pra ~{\bf 99}, 033618 (2019).
		\bibitem{DTC12} C.~Fan, D.~Rossini, H.-X.~Zhang, J.-H.~Wu, M.~Artoni, G.~C.~L.~Rocca, \pra ~{\bf 101}, 013417 (2020).
		\bibitem{DTC13} F.~Machado, D.~V.~Else, G.~D.~K.-Meyer, C.~Nayak, N.~Y.~Yao, Phys.~Rev.~X ~{\bf 10}, 011043 (2020).
		\bibitem{DTC15} P.~Nurwantoro, R.~W.~Bomantara, and J.~Gong, \prb ~{\bf 100}, 214311 (2019). 
		\bibitem{DTC16} K.~Giergiel, A.~Kosior, P.~Hannaford, K.~Sacha, \pra ~{\bf 98}, 013613 (2018). 
		\bibitem{DTC17} K.~Giergiel, A.~Kuros, K.~Sacha, \prb ~{\bf 99}, 220303
		(2019). 
		\bibitem{DTC18} K.~Giergiel, T.~Tran, A.~Zaheer, A.~Singh, A.~Sidorov, K.~Sacha, P.~Hannaford,  New~J.~Phys.~{\bf 22}, 085004 (2020). 
		 \bibitem{DTC19} A.~Kosior, K.~Sacha, \pra ~{\bf 97}, 053621 (2018). 
		 \bibitem{DTC20} P.~Matus, K.~Sacha, \pra ~{\bf 99}, 033626 (2019). 
		 \bibitem{DTC21} A.~Russomanno, S.~Notarnicola, F.~M.~Surace, R.~Fazio, M.~Dalmonte, M.~Heyl,  Phys.~Rev.~Res.~{\bf 2}, 012003 (2020). 
		 \bibitem{DTC22} F.~M.~Surace, A.~Russomanno, M.~Dalmonte, A.~Silva, R.~Fazio, F.~Iemini, \prb ~{\bf 99}, 104303 (2019). 
		 \bibitem{DTC14} K.~Sacha, \emph{Time Crystals} (Springer, Switzerland, 2020).
		 \bibitem{DTC23} D.~V.~Else, C.~Monroe, C.~Nayak, and N.~Y.~Yao, Annu.~Rev.~Condens.~Matter~Phys.~{\bf 11}, 467-499 (2020).
		 \bibitem{DTC24} V.~Khemani, R.~Moessner, S.~L.~Sondhi, arXiv:1910.10745v1.
		 \bibitem{DTC25} K.~Sacha, J.~Zakrzewski, Rep.~Prog.~Phys. ~{\bf 81}, 016401 (2017). 
		 \bibitem{DTC26} F.~Iemini, A.~Russomanno, J.~Keeling, M.~Schiro, M.~Dalmonte, R.~Fazio,  \prl ~{\bf 121}, 035301 (2018). 
		 \bibitem{DTC27} G.~Zlabys, C.-h.~Fan, E.~Anisimovas, K.~Sacha, arXiv:2012.02783. 
		 \bibitem{DTC29} A.~Kuros, R.~Mukherjee, W.~Golletz, F.~Sauvage, K.~Giergiel, F.~Mintert, K.~Sacha, arXiv:2004.14982. 
		 \bibitem{DTC30} J.~Wang, P.~Hannaford, and B.~J.~Dalton, arXiv:2011.14783v1. 
		 \bibitem{DTC31} O.~Shtanko and R.~Movassagh, \prl ~{\bf 125}, 086804 (2020). 
		 \bibitem{DTC32} D.~T.~Liu, J.~Shabani, A.~Mitra, \prb~{\bf 99}, 094303 (2019). 
		\bibitem{DTCqs} A.~Kshetrimayum, M.~Goihl, D.~M.~Kennes, J.~Eisert, arXiv:2004.07267v1.
		\bibitem{DTCqs2} M.~Ippoliti, K.~Kechedzhi, R.~Moessner, S.~L.~Sondhi, and V.~Khemani, arXiv:2007.11602v1. 
		\bibitem{DTCqc} R.~W.~Bomantara and J.~B.~Gong, \prl ~{\bf 120}, 230405 (2018).
		\bibitem{DTCcm1} K.~Sacha, Sci.~Rep.~{\bf 5}, 10787 (2015).
		\bibitem{DTCcm2} D.~Delande, L.~Morales-Molina, and K.~Sacha, \prl ~{\bf 119}, 230404 (2017).
		\bibitem{DTCcm3} K.~Giergiel and K.~Sacha, \pra ~{\bf 95}, 063402 (2017).
		\bibitem{DTCcm5} M.~Mierzejewski, K.~Giergiel, and K.~Sacha, \prb ~{\bf 96},
		140201(R) (2017).
		\bibitem{DTCcm6} K.~Giergiel, A.~Miroszewski, and K.~Sacha, \prl ~{\bf 120}, 140401 (2018).
		(2017).
		\bibitem{FMF1} L.~Jiang, T.~Kitagawa, J.~Alicea, A.~R.~Akhmerov, D.~Pekker, G.~Refael, J.~I.~Cirac, E.~Demler, M.~D.~Lukin, and P.~Zoller, \prl ~{\bf 106}, 220402 (2011). 
		\bibitem{FMF2} D.~E.~Liu, A.~Levchenko, and H.~U.~Baranger, \prl~{\bf 111}, 047002 (2013).
		\bibitem{FMF3} H.-Q.~Wang, M.~N.~Chen, R.~W.~Bomantara, J.~Gong, and D.~Y.~Xing, \prb ~{\bf 95}, 075136 (2017). 
		\bibitem{FMF5} R.~W.~Bomantara and J.~Gong, \prb ~{\bf 98}, 165421 (2018).
		\bibitem{FMF6} Q.-J.~Tong, J.-H.~An, J.~Gong, H.-G.~Luo, and C.~H.~Oh, \prb ~{\bf 87}, 201109(R) (2013).
		\bibitem{FMF7} R.~W.~Bomantara and J.~Gong, J. Phys.: Condens. Matter {\bf 32}, 435301 (2020).
		\bibitem{FMF8} L.~Zhou, \prb ~{\bf 101}, 014306 (2020).
		\bibitem{MBL1} D.~A.~Abanin, W.~De~Roeck, and F.~Huveneers, Ann.~Phys.~(Amsterdam)~{\bf 372}, 1 (2016).
		\bibitem{MBL2} P.~Ponte, A.~Chandran, Z.~Papic, and D.~A.~Abanin, Ann.~Phys.~(Amsterdam)~{\bf 353}, 196 (2015).
		\bibitem{MBL3} P.~Ponte, Z.~Papic, F.~Huveneers, and D.~A.~Abanin, \prl ~{\bf 114}, 140401 (2015).
		\bibitem{MBL4} A.~Lazarides, A.~Das, and R.~Moessner, \prl ~{\bf 115},
		030402 (2015).
		\bibitem{HTI-1} R.-J.~Slager, L.~Rademaker, J.~Zaanen, and L.~Balents, \prb ~{\bf 92}, 085126 (2015). 
	\bibitem{HTI0} W.~A.~Benalcazar, J.~C.~Y.~Teo, and T.~L.~Hughes, \prb ~{\bf 89}, 224503 (2014). 
	\bibitem{HTI1} W.~A.~Benalcazar, B.~A.~Bernevig, and T.~L.~Hughes, Science ~{\bf 357}, 61 (2017). 
	\bibitem{HTI2}  W.~A.~Benalcazar, B.~A.~Bernevig, and T.~L.~Hughes, \prb ~{\bf 96}, 245115 (2017). 
	\bibitem{HTI3} Z.~Song, Z.~Fang, and C.~Fang, \prl ~{\bf 119}, 246402 (2017).
	\bibitem{HTI4} J.~Langbehn, Y.~Peng, L.~Trifunovic, F.~von~Oppen, and P.~W.~Brouwer, \prl ~{\bf 119}, 246401 (2017).
	\bibitem{HTI5} F.~Schindler, A.~M.~Cook, M.~G.~Vergniory, Z.~Wang, S.~S.~P.~Parkin, B.~A.~Bernevig, and T.~Neupert, Sci.~Adv.~{\bf 4}, eaat0346 (2018).
	\bibitem{HTI6} M.~Geier, L.~Trifunovic, M.~Hoskam, and P.~W.~Brouwer, \prb ~{\bf 97}, 205135 (2018).
	\bibitem{HTI7} Z.~Yan, F.~Song, and Z.~Wang, \prl~{\bf 122}, 096803 (2018). 
	\bibitem{HTI8} Q.~Wang, C.~C.~Liu, Y.~M.~Lu, and F.~Zhang, \prl~{\bf 121}, 186801 (2018). 
	\bibitem{HTI9} T.~Liu, J.~J.~He, and F.~Nori, \prb~{\bf 98}, 245413 (2018). 
	\bibitem{HTI10} X.~Zhu, \prb ~{\bf 97}, 205134 (2018).
	\bibitem{HTI11} M.~Ezawa, \prl ~{\bf 120}, 026801 (2018).
	\bibitem{HTI12} E.~Khalaf, \prb ~{\bf 97}, 205136 (2018).
	\bibitem{HTI12b} C.-H.~Hsu, P.~Stano, J.~Klinovaja, and D.~Loss, \prl ~{\bf 121}, 196801 (2018). 
	\bibitem{HTI12c} S.~A.~A.~Ghorashi, X.~Hu, T.~L.~Hughes, E.~Rossi, \prb ~{\bf 100}, 020509(R) (2019). 
	\bibitem{HTI13} F.~K.~Kunst, G.~van~Miert, and E.~J.~Bergholtz, \prb ~{\bf 97}, 241405(R) (2018).
	\bibitem{HTI14} M.~Lin and T.~Hughes, \prb~{\bf 98}, 241103 (2018).
	\bibitem{HTI15} Y.~Xu, R.~Xue, and S.~Wan, arXiv:1711.09202 (2017).
	\bibitem{HTI16} B.~Y.~Xie, H.~F.~Wang, X.~Y.~Zhu, M.~H.~Lu, and Y.~F.~Chen, \prb~{\bf 98}, 205147 (2018).
	\bibitem{HTI17} M.~Serra-Garcia, V.~Peri, R.~S\"{u}sstrunk, O.~R.~Bilal, T.~Larsen, L.~G.~Villanueva, and S.~D.~Huber, Nature (London) ~{\bf 555}, 342 (2018).
	\bibitem{HTI18} F.~Schindler, Z.~Wang, M.~G.~Vergniory, A.~M.~Cook, A.~Murani, S.~Sengupta, A.~Y.~Kasumov, R.~Deblock, S.~Jeon,
	I.~Drozdov, H.~Bouchiat, S.~Guron, A.~Yazdani, B.~A.~Bernevig, and T.~Neupert, Nat.~Phys. ~{\bf 14}, 918-924 (2018).
	\bibitem{HTI19} C.~W.~Peterson, W.~A.~Benalcazar, T.~L.~Hughes, and G.~Bahl, Nature (London) ~{\bf 555}, 346 (2018).
	\bibitem{HTI20} S.~Imhof, C.~Berger, F.~Bayer, J.~Brehm, L.~Molenkamp, T.~Kiessling, F.~Schindler, C.~H.~Lee, M.~Greiter, T.~Neupert,
	and R.~Thomale, Nat.~Phys. ~{\bf 14}, 925-929 (2018). 
	\bibitem{HTI21} L.~Li, M.~Umer, and J.~Gong, \prb~{\bf 98}, 205422 (2018).  
	\bibitem{HTI22} A.~Matsugatani and H.~Watanabe, \prb~{\bf 98}, 205129 (2018).
	\bibitem{HTI23} S.~Franca, J.~van~den~Brink, and I.~C.~Fulga, \prb~{\bf 98}, 201114 (2018). 
	\bibitem{HTI24} J.~Noh, W.~A.~Benalcazar, S.~Huang, M.~J.~Collins, K.~P.~Chen, T.~L.~Hughes, and M.~C.~Rechtsman, Nat.~Photon. ~{\bf 12}, 408-415 (2018). 
	\bibitem{HTI25} H.~Xue, Y.~Yang, F.~Gao, Y.~Chong, and B.~Zhang, Nat.~Mater.~{\bf 18}, 108-112 (2019). 
	\bibitem{HTI26} D.~Calugaru, V.~Jurici\'{c}, and B.~Roy, Phys.~Rev.~B {\bf 99}, 041301(R) (2019). 
	\bibitem{FHTI1} B.~Huang and W.~V.~Liu, arXiv:1811.00555. 
	\bibitem{FHTI2} R.~W.~Bomantara, L.~Zhou, J.~Pan, and J.~Gong, \prb ~{\bf 99}, 045441 (2019). 
	\bibitem{FHTI3} M.~R.-Vega, A.~Kumar, and B.~Seradjeh, \prb ~{\bf 100}, 085138 (2019). 
	\bibitem{FHTI4} H.~Hu, B.~Huang, E.~Zhao, and W.~V.~Liu, \prl ~{\bf 124}, 057001 (2020). 
	\bibitem{FHTI5} R.~Seshadri, A.~Dutta, and D.~Sen, \prb ~{\bf 100}, 115403 (2019).
	\bibitem{FHTI6} K.~Plekhanov, M.~Thakurathi, D.~Loss, and J.~Klinovaja, Phys.~Rev.~Research {\bf 1}, 032013(R) (2019). 
	\bibitem{FHTI7} T.~Nag, V.~Jurici\'{c}, and B.~Roy,
	Phys.~Rev.~Research {\bf 1}, 032045(R) (2019). 
	\bibitem{FHTI8} R.~W.~Bomantara and J.~Gong, \prb ~{\bf 101}, 085401 (2020).
	\bibitem{FHTI9} R.~W.~Bomantara, Phys.~Rev.~Research~{\bf 2}, 033495 (2020)
		\bibitem{Flo1} J.~H.~Shirley, Phys.~Rev.~{\bf 138}, B979 (1965).
		\bibitem{Flo2} H.~Sambe, \pra ~{\bf 7}, 2203 (1973).
		\bibitem{Chapman} A.~Chapman and S.~T.~Flammia, Quantum~{\bf 4}, 278 (2020). 
		\bibitem{pzmandppm} G.~J.~Sreejith, A.~Lazarides, and R.~Moessner, \prb ~{\bf 94}, 045127 (2016).	\bibitem{Zhangexp} W.~Zhu, H.~Xue, J.~Gong, Y.~Chong, B.~ Zhang, arXiv:2012.08847. 
		\bibitem{DTCexp1} J.~Zhang, P.~W.~Hess, A.~Kyprianidis, P.~Becker, A.~Lee, J.~Smith, G.~Pagano, I.-D.~Potirniche, A.~C.~Potter, A.~Vishwanath, N.~Y.~Yao, and C.~Monroe, Nature (London)~{\bf 543}, 217 (2017).
		\bibitem{DTCexp7} A.~Kyprianidis, F.~Machado, W.~Morong, P.~Becker, K.~S.~Collins, D.~V.~Else, L.~Feng, P.~W.~Hess, C.~Nayak, G.~Pagano, N.~Y.~Yao, and C.~Monroe, arXiv:2102.01695v1
		 \bibitem{DTCexp2} S.~Choi, J.~Choi, R.~Landig, G.~Kucsko, H.~Zhou, J.~Isoya, F.~Jelezko, S.~Onoda, H.~Sumiya, V.~Khemani, C.~v.~Keyserlingk, N.~Y.~Yao, E.~Demler, and M.~D.~Lukin, Nature (London)~{\bf 543},
		 221 (2017).
		 \bibitem{DTCexp3} J.~Rovny, R.~L.~Blum, and S.~E.~Barrett, \prl~{\bf 120},
		 180603 (2018).
		 \bibitem{DTCexp4} J.~Rovny, R.~L.~Blum, and S.~E.~Barrett, \prb~{\bf 97}, 184301
		 (2018).
		 \bibitem{DTCexp5} S.~Pal, N.~Nishad, T.~S.~Mahesh, and G.~J.~Sreejith, \prl ~{\bf 120}, 180602 (2018).
		\bibitem{Sycamore} F.~Arute \emph{et al}, Nature {\bf 574}, 505 (2019).
		\bibitem{repDTC} R.~W.~Bomantara, arXiv:2102.09113.
	\end{thebibliography}

\end{document}